\documentclass[aps,prb,twocolumn,groupedaddress,showpacs, citeautoscript,superscriptaddress]{revtex4-2}
\usepackage{graphicx}
\usepackage{dcolumn}
\usepackage{bm}
\usepackage{amsmath,amsfonts,amssymb,amsthm,verbatim}
\usepackage[ansinew]{inputenc}
\usepackage{color}

\usepackage{epsfig}
\usepackage{wasysym}

\begin{document}

\title{Detecting and leaking a Majorana bound state through proximity to a Kitaev ring}

\author{Mariana Malard}
\email{mmalard@unb.br}
\affiliation{ Faculdade UnB Planaltina and International Center of Physics, Universidade de Bras\'{i}lia, Bras\'{i}lia-DF, Brazil}

\author{David S. Brand\~ao }

\affiliation{ Department of Physics, University at Buffalo, State University of New York, Buffalo, New York 14260, USA}

\date{\today}
\begin{abstract}
We show that the existence of a Majorana bound state at one end of a Kitaev chain is unambiguously signaled by observable quantities in a nearby Kitaev ring. When the Kitaev chain is in the topological phase, the band structure of the Kitaev ring breaks chiral symmetry and the ring-Majoranas' spectral functions are even functions of momentum and energy, the latter wielding time-reversal and particle-hole symmetric occupation numbers and, hence, zero current in the ring. Driving a phase transition in the Kitaev chain (e.g., using a backgate to vary the chemical potential of the chain through its critical value), the replacement of the Majorana bound state by a trivial fermion is manifest by chiral symmetry reappearing in the band structure of the ring and by the ring-Majoranas' spectral functions loss of parity, wielding time-reversal and particle-hole symmetry breaking in the occupation numbers and a spontaneous current in the ring. Energy resonance and maximum spectral weight exactly at the resonance energy concur for a high leakage probability between ring-Majoranas' states and the Majorana bound state, whereas the lack of energy resonance makes unlikely any leakage between ring-Majoranas' states and the trivial fermion states. The proposed setup also invites further investigations in the context of braiding and/or fusion of Majorana states, both as detection mechanisms and for quantum computing purposes.
\end{abstract}

\maketitle

\section{Introduction}
\label{sec:intro}
It has now been more than two decades of Kitaev's model of a one-dimensional topological superconductor hosting zero-energy Majorana bound states (MBSs) \cite{Kitaev2001}. Since then, a theoretical and experimental tour de force has been in place to realize Kitaev's chain as the detection of a MBS has formidable implications for physics and technological applications. Indeed, while Majorana fermions were first predicted 86 years ago \cite{Majorana1937} and remain unobserved as fundamental particles, braiding Majorana states in topological superconductors could lead up to a topological $q$-bit \cite{Kitaev2003,Nayak2008,Bonderson2008,Alicea2011,Heck2012,Vijay2016,Karzig2017,Beenakker2019}. But, defying all efforts, Majorana particles remain elusive in free space and in matter. The zero-bias conductance peak, prominently featured in well-known attempts to realize Kitaev's chain \cite{Lutchyn2010,Oreg2010} and therein considered as a signature of MBSs, is, in fact, ubiquitous to a variety of transport measurements, appearing in quantum dots due to the Kondo effect \cite{Gordon1998,Cronenwett1998}, in metal-superconductor junctions from Andreev bound states \cite{Golubov2009}, and in a range of disordered systems \cite{Mourik2012,Deng2012,Das2012,Lee2012,Finck2013,Law2009,Pikulin2012,Churchill2013}. Alternative experimental signatures of MBSs are thus in high demand, a promising direction aiming at their distinctive non-Abelian statistics \cite{Shabani2016,Fatin2016,Matos-Abiague2017,Pientka2017,Hell2017,Scharf2019,Liu2019,Zhou2019,Setiawan2019,Hedge2020,Alidoust2021,Zhou2022}.

Another strategy is to use an ancillary system to probe the presence of a nearby MBS. It has been proposed that the conductance through a quantum dot coupled to a MBS signals its presence \cite{Liu2011,Vernek2014,Smirnov2015}. Combining quantum dots and MBSs leads to interference, which could also yield a detection mechanism \cite{Gong2014,Zambrano2018}. Although theoretically appealing, these proposals have not yet thrived in the experimental arena since, arguably, the predicted MBS signature is likely to be spoiled by decoherence. In another theoretical proposal, the cancellation of the persistent current in a ring pierced by a magnetic flux is attributed to the presence of nearby MBSs \cite{Gong2015}. Related ring-geometries were explored to investigate interference in the conductance \cite{Chiu2018}, interference contrast \cite{Hell2018}, and the change in the MBSs parity induced by the magnetic flux \cite{Medina2020}. Although persistent current setups may become useful as practical detection tools for MBSs, the reliance on a magnetic field hinders their utility in quantum computing designs. Indeed, a universal set of quantum gates \cite{Nielsen2011} requires auxiliary nontopological states \cite{Bravyi2005,Bravyi2006} which are known to be fragile to magnetic field.

In this paper we propose a simple, magnetic-field free, setup consisting of a ``Kitaev ring" - a closed Kitaev chain - coupled to one of the MBSs of a Kitaev open chain in the topological phase, c.f. Fig. 1(a). Approaching the chain and the ring will cause the wave functions of the ring-Majoranas and the nearby MBS to start overlapping, with the result that the latter may leak into one of the ring states or vice-versa. In either case, the ring is left with a delocalized Majorana state, unlike the original MBS which was pinned to the edge of the chain. To draw relevance to the longstanding conundrum of MBS \emph{versus} trivial bound states (TBSs), we also analyze the case in which the ring is coupled to a fermionic TBS whose energy is so small that said TBS could pass as a MBS in measurements of zero-bias conductance or in local probes of the edge. We find that the band structures, spectral functions, occupation numbers and currents in the MBS and TBS pictures respond to the edge-ring coupling in fundamentally different ways.

The current in the ring offers a particularly sharp distinction: While it vanishes identically in the MBS picture, the TBS picture supports different particle and hole currents which appear spontaneously (in that the ring is not subject to a voltage, nor is it thread by a magnetic field) from an unbalance in those carriers' occupation numbers. This unbalance stems from the ring-TBS coupling breaking particle-hole and time-reversal symmetries of the occupation numbers, although those symmetries are preserved in the band structure. This situation partly resembles the appearance of an asymmetric conductance with respect to positive and negative voltage which has been reported in a wealth of proposed realizations of particle-hole asymmetric topological superconductors, spanning over twenty years \cite{Yazdani1997,Matsuba2003,Shan2011,Hanaguri2012,Suominen2017,Nichele2017,Menard2017,Choi2017,Gul2018,Deng2018,Chen2018,Bommer2019,Chen2019,Vaitiekenas2020,Saldana2020,Farinacci2020,Yu2021,Wang2021,Ding2021}. The breaking of particle-hole symmetry in such systems has been attributed to ``quasi-particle poisoning" by which Andreev bound states decay into the continuum of quasiparticles of the superconductor \cite{Ruby2015} or are coupled to fermionic \cite{Martin2014,DasSarma2016,Liu2017} or bosonic \cite{Setiawan2021} baths. Despite their similarities, those previous and the present works strikingly differ in that while the current is driven by a voltage therein, here it is spontaneous. Persistent currents at zero voltage and zero magnetic field have been predicted to arise in, e.g., a graphene ring coupled to two reservoirs of particles \cite{Benjamin2014}, and more generically conceptualized within the framework of the nonreciprocal Lindblad dynamics, where the non-reciprocity is generated by a structured dissipative environment \cite{Keck2018}, causing time-reversal symmetry breaking. This scenario indicates that coupling to external modes may induce changes in the electric transport due to particle-hole symmetry breaking and spontaneous electric currents from time-reversal symmetry breaking. Lying in the intersection between these two phenomena, the setup proposed here provides experimentally accessible signatures of the coupling of a Kitaev ring to a MBS \emph{or} to a TBS and thus may be used as an unambiguous MBS detector.

Moreover, for its geometry and the possibility of leakage between ring-Majoranas' states and the MBS, the proposed setup also invites the investigation of braiding and/or fusion between these states, with possible relevant implications for MBS detection and quantum computing.

\section{Models and spectra}

The Hamiltonian which describes the setup depicted in Fig. 1(a) is $H=H_{\text{e}}+H_{\text{r}}+H_{\text{er}}$ with
\begin{equation}\label{He}
H_{\text{e}}=\varepsilon_{\text{e}}\,\gamma_{\text{e}}^{\dagger}\gamma_{\text{e}},
\end{equation}
\begin{equation}\label{Hr}
H_{\text{r}}=\sum_{j=1}^{N}\left[-t\,c_{j}^{\dagger}c_{j+1}+\mu\,c_{j}^{\dagger}c_{j}-\Delta\,c_{j}^{\dagger}c_{j+1}^{\dagger}+H.c.\right],
\end{equation}
\begin{equation}\label{Her}
H_{\text{er}}=\sum_{j=1}^{N}\sum_{\tau=1,2}[v_{\tau}(j)\,\gamma_{\tau,j}^{\dagger}\gamma_{\text{e}}+H.c.],
\end{equation}
where $H_{\text{e}}$ is the ``edge-Hamiltonian" describing a MBS with energy $\varepsilon_{\text{e}}=0$ (but which we keep as a generic parameter for now), with $\gamma_{\text{e}}^{\dagger}=\gamma_{\text{e}}$ the Majorana operator; $H_{\text{r}}$ is the usual Hamiltonian of a Kitaev chain with $N$ sites in which spinless fermions, represented by the fermionic operator $c_{j}$ at site $j$, hop with amplitude $t$, have on-site energy $\mu$ and are paired with superconducting amplitude $\Delta$ (with the difference with respect to the Kitaev chain that here it is closed in a ring and, hence, periodic boundary condition is automatically implied); and $H_{\text{er}}$ is the Hamiltonian which couples the MBS $\gamma_{\text{e}}$ and the ring-Majoranas $\gamma_{\tau,j}^{\dagger}=\gamma_{\tau,j}$ through the hopping amplitudes $v_{\tau}(j)$. The Majorana and fermionic operators in the ring are related by: $\gamma_{\tau,j}=(i)^{\tau-1}\,[\,c_{j}^{\dagger}+(-1)^{\tau-1}c_{j}\,]$.
\begin{figure}
\begin{center}
\includegraphics[width=\linewidth]{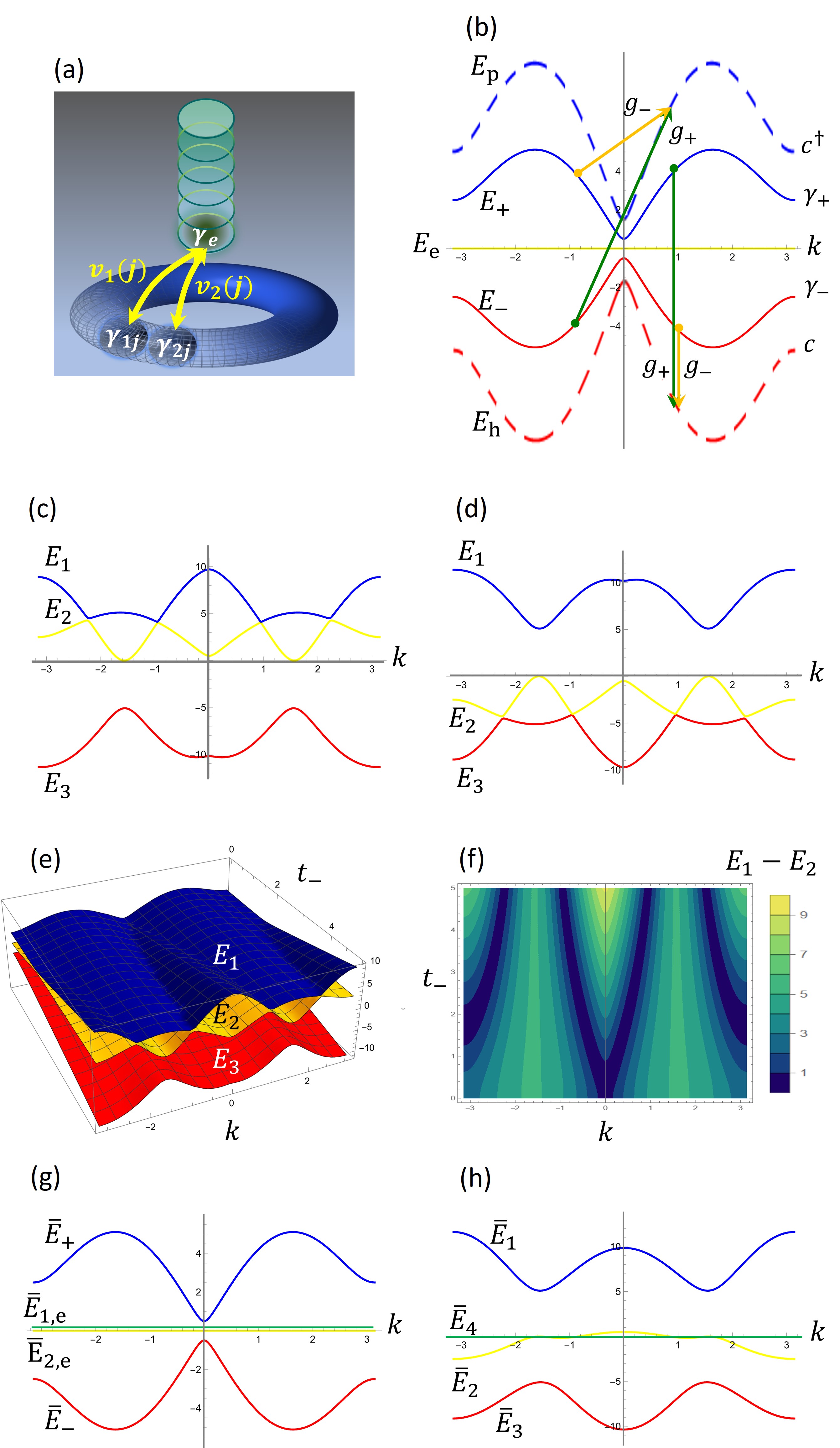}
\caption{(a) A Kitaev ring, with its pair of ring-Majoranas $\gamma_{1/2,j}$ at site $j$, is coupled to one of the Majorana bound states $\gamma_{\text{e}}$ of a long Kitaev chain through couplings $v_{1/2}(j)$. (b) Bands $E_{\pm}$ of the ring-Majoranas and $E_{\text{e}}$ of the Majorana bound state as a function of wave number $k$, with $t=1.5$, $\mu=2$, and $\Delta=5$ (putting the ring in the gapped topological phase), $t_{+}=t_{-}=\mu_{+}=\mu_{-}=\Delta_{+}=\Delta_{-}=0$ (that is, for zero coupling between the ring and the Majorana bound state), and $\varepsilon_{\text{e}}=0$ (which is the energy of the Majorana bound state). $E_{\text{p}/\text{h}}$ are the bands of particles and holes in this uncoupled ring. The relations between ring-Majoranas' states and particles and holes' states are illustrated by the factors $g_{\pm}$. (c) [(d)] Bands $E_{1-3}$ of the hybridized states, with $t=1.5$, $\mu=2$, $\Delta=5$, $t_{+}=0.1\,[5]$, $t_{-}=5\,[0.1]$, $\mu_{+}=\mu_{-}=\Delta_{+}=\Delta_{-}=0.05$, and $\varepsilon_{\text{e}}=0$. (e)-(f) Bands $E_{1-3}$ (in (e)) and color map of the gap between bands $E_{1}$ and $E_{2}$ (in (f)) as a function of $k$ and $t_{-}$, with the other coupling parameters as in (c). (g) Bands $\bar{E}_{\pm}$ of the ring-Majoranas and $\bar{E}_{1/2,\text{e}}$ of the edge-Majoranas which pair up to form a trivial bound state as a function of wave number $k$, with $t=1.5$, $\mu=2$, $\Delta=5$ (putting the ring in the gapped topological phase), $\bar{t}=\bar{\mu}=\bar{\Delta}=0$ (that is, for zero coupling between the ring and the trivial bound state), and $\bar{\varepsilon}_{\text{e}}=0$ (for a putative zero-energy trivial bound state). (h) Bands $\bar{E}_{1-4}$ of the hybridized states, with $t=1.5$, $\mu=2$, $\Delta=5$, $\bar{t}=10.2$, $\bar{\mu}=\bar{\Delta}=0.2$, and $\bar{\varepsilon}_{\text{e}}=0$.}
\label{Fig1}
\end{center}
\end{figure}

Fourier transforming and changing to the diagonal basis of $H_{\text{r}}$, the full Hamiltonian $H$ reads
\begin{equation}\label{FTH}
H=\sum_{k}\psi_{k}^{\dagger}\,{\cal H}(k)\,\psi(k),
\end{equation}
where $k=2\pi m/N$ with $m=0,\pm1,...,\pm N/2$ ($k/a$, with $a$ the lattice spacing, are the wave numbers in the Brillouin zone), $\psi_{k}^{\dagger}=[\gamma_{+,k}^{\dagger}\,\,\,\gamma_{-,k}^{\dagger}\,\,\,\gamma_{\text{e}}^{\dagger}]$, $\gamma_{\pm,k}=(1/\sqrt{2})[\,(f^{\ast}(k)/|f(k)|)\,\gamma_{1,k}\,\pm\,\gamma_{2,k}\,]$, and the $3\times3$ Bloch Hamiltonian
\begin{equation}\label{BlochH}
{\cal H}(k)=\begin{bmatrix}
    |f(k)| & 0 & \omega_{+}(k)\\
    0 & -|f(k)| & \omega_{-}(k)\\
    \omega_{+}^{\ast}(k) & \omega_{-}^{\ast}(k) & \varepsilon_{\text{e}}/N\\
\end{bmatrix},
\end{equation}
with $f(k)=0.5\,[\,2\Delta\sin k\,+\,i(-2t\cos k+\mu)\,]$ and $\omega_{\pm}(k)=(1/\sqrt{2})\,[\,(f^{\ast}(k)/|f(k)|)\,v_{1}(k)\,\pm\,v_{2}(k)\,]$. The relation between the $k$-space operators $\gamma_{\pm,k}$ and $c_{k}$ is: $c_{k}=g_{+}(k)\gamma_{+,k}+g_{-}(k)\gamma_{-,k}$, with $g_{\pm}(k)=(1/2\sqrt{2})\,[\,(f(k)/|f(k)|)\,\pm\,i\,]$.

We note that $H_{\text{r}}$, whose $2\times2$ Bloch Hamiltonian is simply ${\cal H}_{\text{r}}(k)=\text{diag}[|f(k)| \,\,\, -|f(k)|]$, has both time-reversal ($T$) and chiral ($S$) symmetries. The full Hamiltonian $H$ in Eq. (\ref{FTH}) preserves $T$ if the entries in Eq. (\ref{BlochH}) obey $\omega_{\pm}^{\ast}(-k)=\omega_{\pm}(k)$, implying $\omega_{\pm}(k)$ of the form $\omega_{\pm}(k)=(-2t_{\pm}\cos k+\mu_{\pm})\,-\,i(2\Delta_{\pm}\sin k)$. However, together these off-diagonal entries cause $H$ to break $S$.

As a base for comparison, Fig. 1(b) shows the bands of ${\cal H}(k)$ given in Eq. (\ref{BlochH}), for zero ring-MBS coupling, that is, for $\omega_{\pm}(k)\equiv0$, with $\varepsilon_{\text{e}}=0$ and a choice of parameters which puts the ring in the gapped topological phase (defined by $t>\mu/2$ for $\mu>0$). The ring-Majoranas' bands $E_{\pm}(k)=\pm|f(k)|$ (solid blue and solid red) are $T$- and $S$-symmetric, and $E_{\text{e}}=0$ (solid yellow) is the band of the MBS, which is just the expected result for the uncoupled case. Bands $E_{\text{p}/\text{h}}(k)$ (dashed blue and dashed red) are those of the fermionic particles and holes in the uncoupled ring. The green and orange arrows, tagged with the $g_{\pm}(k)$-factors, indicate the relations between the fermionic and Majorana states in the uncoupled ring. Turning on the coupling induces hybridization between the ring-Majoranas and the MBS. Figs. 1(c)-(d) display the $T$-symmetric, but now $S$-nonsymmetric, bands $E_{1}(k)$ (blue), $E_{2}(k)$ (yellow), and $E_{3}(k)$ (red) of the hybridized states for two choices of ring-MBS coupling parameters. $E_{2}(k)$ touches $E_{1/3}(k)$ in Fig. 1(c)/(d) at two pairs of $T$-symmetric momenta in the Brillouin zone for that choice of parameters. Varying one of the coupling parameters makes these band-touching points move in the Brillouin zone, as can be seen in Figs. 1(e)-(f) by varying $t_{-}$.

We now compare the previous band structures to those of the case where the Kitaev ring is coupled to a fermionic TBS. The Hamiltonian is now $\bar{H}=\bar{H}_{\text{e}}+H_{\text{r}}+\bar{H}_{\text{er}}$, with the same $H_{\text{r}}$ as in Eq. (\ref{Hr}) and
\begin{equation}\label{Hebar}
\bar{H}_{\text{e}}=\bar{\varepsilon}_{\text{e}}\,c_{\text{e}}^{\dagger}c_{\text{e}},
\end{equation}
\begin{equation}\label{Herbar}
\bar{H}_{\text{er}}=\sum_{j=1}^{N}[\bar{v}(j)\,c_{j}^{\dagger}c_{\text{e}}+H.c.],
\end{equation}
being, respectively, the Hamiltonian of the TBS with energy $\bar{\varepsilon}_{\text{e}}$ and fermionic operator $c_{\text{e}}$, and the Hamiltonian which couples the TBS and ring-fermions through the hopping $\bar{v}(j)$. The Majorana operators for the TBS and the ring read $\gamma_{\tau,\xi}=(i)^{\tau-1}\,[\,c_{\xi}^{\dagger}+(-1)^{\tau-1}c_{\xi}\,]$ with $\tau=1,2$ and $\xi=\text{e},j$.

Again Fourier transforming and changing to the diagonal basis of $H_{\text{r}}$, we now obtain
\begin{equation}\label{FTHbar}
\bar{H}=\sum_{k}\bar{\psi}_{k}^{\dagger}\,{\cal \bar{H}}(k)\,\bar{\psi}(k),
\end{equation}
where $\bar{\psi}_{k}^{\dagger}=[\gamma_{+,k}^{\dagger}\,\,\,\gamma_{-,k}^{\dagger}\,\,\,\gamma_{1,\text{e}}^{\dagger}\,\,\,\gamma_{2,\text{e}}^{\dagger}]$, and the $4\times4$ Bloch Hamiltonian
\begin{equation}\label{BlochHbar}
{\cal \bar{H}}(k)=\begin{bmatrix}
    |f(k)| & 0 & u_{+,1}(k) & u_{+,2}(k)\\
    0 & -|f(k)| & u_{-,1}(k) & u_{-,2}(k)\\
    u_{+,1}^{\ast}(k) & u_{-,1}^{\ast}(k) & \bar{\varepsilon}_{\text{e}}/4N & i\bar{\varepsilon}_{\text{e}}/4N\\
    u_{+,2}^{\ast}(k) & u_{-,2}^{\ast}(k) & -i\bar{\varepsilon}_{\text{e}}/4N & \bar{\varepsilon}_{\text{e}}/4N\\
\end{bmatrix},
\end{equation}
with $u_{\pm,\tau}(k)=0.5\,(i)^{\tau-1}\,\bar{v}(k)\,g_{\pm}^{\ast}(k)$. $\bar{H}$ preserves $T$ if $\bar{v}^{\ast}(-k)=\bar{v}(k)$, which implies $\bar{v}(k)=(-2\bar{t}\cos k+\bar{\mu})\,-\,i(2\bar{\Delta}\sin k)$. Since $g_{\pm}(-k)=-g_{\pm}^{\ast}(k)$, one gets $u_{\pm,\tau}^{\ast}(-k)=(-1)^{\tau}\,u_{\pm,\tau}(k)$.

To consider the scenario of a TBS disguising as a MBS we take $\bar{\varepsilon}_{\text{e}}=0$, in which case one of the bands of ${\cal \bar{H}}(k)$ in Eq. (\ref{BlochHbar}) is $\bar{E}_{1,\text{e}}=0$ and the others are computed from the reduced Hamiltonian
\begin{equation}\label{BlochHbar3bands}
{\cal \bar{H}}_{\text{red}}(k)=\begin{bmatrix}
    |f(k)| & 0 & \bar{\omega}_{+}(k)\\
    0 & -|f(k)| & \bar{\omega}_{-}(k)\\
    \bar{\omega}_{+}^{\ast}(k) & \bar{\omega}_{-}^{\ast}(k) & 0\\
\end{bmatrix},
\end{equation}
where $\bar{\omega}_{\pm}(k)=u_{\pm,1}(k)+u_{\pm,2}(k)$. A superficial comparison of Eqs. (\ref{BlochH}) and (\ref{BlochHbar3bands}) could mislead one to conclude that, when $\bar{\omega}_{\pm}(k)=\omega_{\pm}(k)$, the MBS picture (with $\varepsilon_{\text{e}}=0$, which is the physical value) and the TBS picture (with putative $\bar{\varepsilon}_{\text{e}}=0$) yield three identical bands. However, $\bar{\omega}_{\pm}(k)$ and $\omega_{\pm}(k)$ cannot, in general, be chosen to coincide because while $\omega_{+}(k)$ and $\omega_{-}(k)$ can be taken independently via $v_{1}(k)$ and $v_{2}(k)$, $\bar{\omega}_{+}(k)$ and $\bar{\omega}_{-}(k)$ are both constrained to the single coupling $\bar{v}(k)$. As a matter of fact, this constraint (and the relation between $\bar{\omega}_{\pm}(k)$ and $g_{\pm}(k)$) ensures $S$ in half of the bands of $\bar{H}$, which $H$ lacks entirely.

The bands of ${\cal\bar{H}}_{\text{red}}(k)$, as given in Eq. (\ref{BlochHbar3bands}), plus the band $\bar{E}_{1,\text{e}}=0$, are shown in Fig. 1(g) for zero ring-TBS coupling, with the same choice of ring-parameters as in Fig. 1(b) so that $\bar{E}_{\pm}(k)=E_{\pm}(k)$ (blue and red). $\bar{E}_{1,\text{e}}=\bar{E}_{2,\text{e}}=0$ (green and yellow) are the doubly-degenerate bands of the two edge-Majoranas which form the TBS. Fig. 1(h) shows the bands when $\bar{v}(k)=2[\omega_{+}(k)+\omega_{-}(k)]$, with the same parameters in $\omega_{\pm}(k)$ as in Fig. 1(c). This relation between the coupling in the TBS picture, $\bar{v}(k)$, and those in the MBS picture, $\omega_{\pm}(k)$, sets the same band width in Figs. 1(c) and (h). The hybridized bands $\bar{E}_{1}(k)$ (blue) and $\bar{E}_{3}(k)$ (red) in Fig. 1(h) are $T$- and $S$-symmetric, while $S$ is broken between the hybridized bands $\bar{E}_{4}=0$ (green) and $\bar{E}_{2}(k)$ (yellow). Finally, there is no gap closing between $\bar{E}_{2}(k)$, which disperses close to zero energy, and either $\bar{E}_{1}(k)$ or $\bar{E}_{3}(k)$.

In sum, the band structures of the MBS and the zero-energy TBS pictures differ in two fundamental ways: (i) The former has only one MBS band at zero energy which, upon hybridization due to the coupling with the ring, is fully pushed above or below zero energy and touches one of the other two bands for given values of the ring-MBS couplings. Conceivably, at the momenta where the gap vanishes, there might be leakage between the equal-energy hybridized states. Meanwhile, in the zero-energy TBS picture, the band structure has two edge-Majorana bands at zero energy at zero coupling, with one hybridized band sticking to zero energy and the other dispersing around zero, but far from the other bands for ring-TBS couplings energetically consistent with those of the MBS picture. The lack of energy resonance makes leakage unlikely between the hybridized states. (ii) Unlike the bands of the MBS picture which, at nonzero ring-MBS coupling, break $S$ of the uncoupled ring, two bands of the TBS picture retain that symmetry at nonzero ring-TBS coupling. 

\section{Spectral functions, occupation numbers and currents}

The Green's functions $G_{\nu}(k,\omega+i\eta)\equiv\,\ll\gamma_{\nu,k}:\gamma_{\nu,k}^{\dagger}\gg_{\omega+i\eta}$, $\nu$ encoding the particle species, are computed from their equations of motion and Dyson equation (c.f. Appendix A). The imaginary parts of the Green's functions of the ring-Majoranas read in both pictures
\begin{equation}\label{ImMain}
\text{Im}\,G_{\pm}(k,\omega)=\frac{0.25\Lambda(k,\omega)}{[\omega\mp|f(k)|-0.5M(k,\omega)]^{2}+[0.5\Lambda(k,\omega)]^{2}},
\end{equation}
where $M[\Lambda]\,=\,\text{Lim}_{\eta\rightarrow0}\,\text{Re[Im]}\,\Sigma$ and $\Sigma$ the self-energy.

In the MBS picture, the imaginary part of the $k$-independent MBS Green's function is
\begin{equation}\label{ImeMain}
\text{Im}\,G_{\text{e}}(\omega)=\frac{2G(\omega)}{F^{2}(\omega)+G^{2}(\omega)},
\end{equation}
where $F[G]\,=\,\text{Lim}_{\eta\rightarrow0}\,\text{Re[Im]}\,\Sigma_{\text{e}}$ and $\Sigma_{\text{e}}$ the MBS self-energy.

The spectral functions of the ring-Majoranas $\gamma_{\pm,k}$ in both pictures and of the MBS $\gamma_{\text{e}}$ in the MBS picture are given by: $A_{\nu}=-\text{Im}\,G_{\nu}$, $\nu=\pm,\text{e}$. Eq. (\ref{ImMain}) with $M(k,\omega)=\Lambda(k,\omega)=0$ yields that an uncoupled ring has ring-Majoranas' spectral functions $A_{\pm}^{(0)}(k,\omega)=0.5\delta(\omega-E_{\pm}(k))$, where $E_{\pm}(k)=\pm|f(k)|$ are the unperturbed non-hybridized ring-Majoranas' bands (c.f. Fig. 1(b)). Eq. (\ref{ImMain}) thus implies that coupling the ring to the edge turns $A_{\pm}(k,\omega)$ from Dirac delta functions centered at $E_{\pm}(k)$ into Lorentzian ones which are shifted from $E_{\pm}(k)$ by a factor $0.5M(k,\omega)$, entailing the correction to the ring-Majoranas' energies due to the coupling. Also, the spectral peaks acquire a width $0.5\Lambda(k,\omega)$, wielding a nonzero lifetime to the ring-Majoranas, and height $[0.5\Lambda(k,\omega)]^{-1}$, encoding the ring-Majoranas' now finite spectral weight. In the case of an uncoupled MBS, Eq. (\ref{ImeMain}) with $F(\omega)=G(\omega)=0$ yields $A_{\text{e}}^{(0)}(\omega)=2\delta(\omega)$. Turning on the ring-MBS coupling causes the spectral weight of the now finite-living MBS to distribute over an energy range according to Eq. (\ref{ImeMain}).

Figs. 2(a)-(b) show a color map of $A_{\pm}(k,\omega)$ in the MBS (panel (a)) and TBS (panel (b)) pictures (with $E_{\pm}(k)$ shown in yellow for reference). The plots convey that, for a choice of parameters equivalent to that of Figs. 1 (c) and (h), the spectral peaks of the ring-Majoranas are higher at energies $\omega\sim\pm5\,\text{a.u}$ and wave numbers $k\sim\pm0.5\pi$ (in units of lattice spacing) \emph{in both pictures}. Fig. 2(c), with same choice of parameters, shows that $A_{\text{e}}(\omega)$ changes from the Dirac delta function centered at zero energy (depicted in red) to peaks away from zero energy when the ring-MBS coupling is turned on. The maximum spectral weight of the MBS at $\omega\sim\pm5\,\text{a.u}$ coincides with the maximum spectral weights of the ring-Majoranas in Fig. 1(a), with $\omega\sim+5\,\text{a.u}$ being the resonance energy between bands $E_{2}(k)$ and $E_{1}(k)$  (c.f. Fig. 1(c)). Choosing parameters equivalent to those in Fig. 1(d) yields similar ring-Majoranas' and MBS spectral functions, with the energy resonance now occurring between $E_{2}(k)$ and $E_{3}(k)$ (c.f. Fig. 1(d)). The energy resonance between the hybridized states combined with the maximum in the ring-Majoranas' and MBS spectral functions exactly at the resonance energy concur for a high leakage probability between the resonating states. Differently, since in the TBS picture $\omega\sim\pm5\,\text{a.u}$ do not correspond to resonance energies between any of the bands (c.f. Fig. 1(h)), one does not expect an enhanced leakage probability between the states at those energies, even though those energies also maximize the states' spectral weight in the TBS picture.

Figs. 2(a)-(b) also convey that, despite the occurrence of the maxima of the ring-Majoranas' spectral functions at the same energy and momenta in the MBS and TBS pictures, in the MBS picture the ring-Majoranas' spectral functions obey the parity relations $A_{\pm}(-k,\omega)=A_{\pm}(k,\omega)$ and $A_{-}(k,-\omega)=A_{+}(k,\omega)$, while these relations are lacking in the TBS picture. Fig. 3 shows that the extent of the parity breaking of $A_{\pm}(k,\omega)$ in the TBS picture depends on the value of $\omega$. This striking difference between the MBS and TBS pictures is related to the coupling between the ring-Majoranas to only one Majorana edge-state in the MBS picture \emph{versus} the coupling to two Majorana edge-states in the TBS picture, the latter generating parity-breaking terms in the ring-Majoranas' self-energies (c.f. Appendix A). Formally, the difference can be traced to the fact that the single edge-Majorana operator ($\gamma_{\text{e}}$) commutes with the edge-Hamiltonian ($H_{\text{e}}$) in the MBS picture while, in the TBS picture, the edge-Majorana operators ($\gamma_{1,\text{e}}$ and $\gamma_{2,\text{e}}$) do not individually commute with the edge-Hamiltonian ($\bar{H}_{\text{e}}$).
\begin{figure}
\begin{center}
\includegraphics[width=\linewidth]{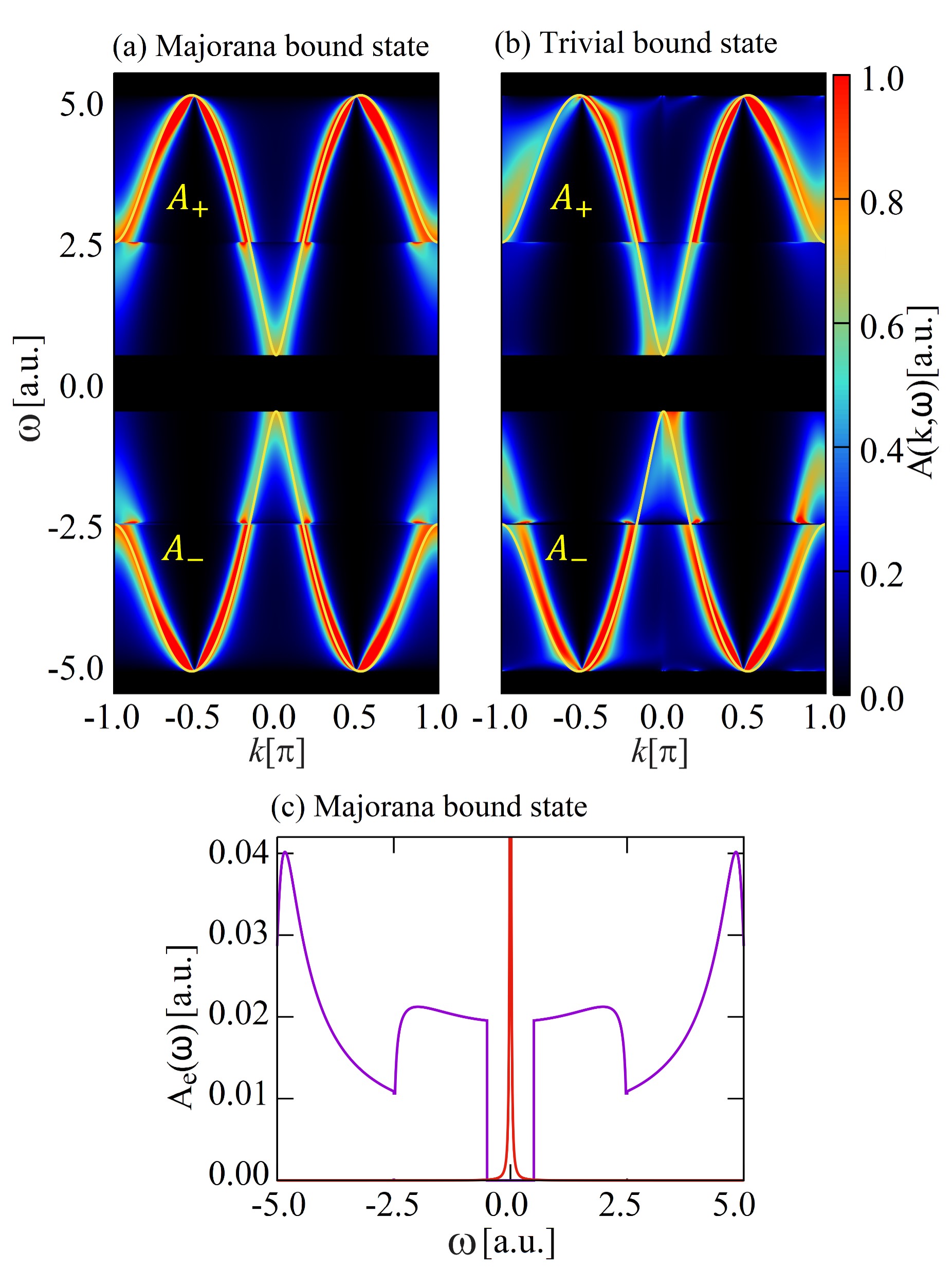}
\caption{(a) [(b)] Color map of the spectral functions $A_{\pm}$ of the ring-Majoranas as a function of wave number $k$ (in units of lattice spacing) and energy $\omega$ in the Majorana bound state [trivial bound state] picture. The yellow lines are the bands of the ring-Majoranas for an uncoupled ring. (c) Spectral function $A_{\text{e}}$ of the Majorana bound state as a function of energy $\omega$ in the Majorana bound state picture. The red peak illustrates the spectral function of the Majorana bound state when it is uncoupled - a Dirac delta function centered at zero energy. In all plots, $t=1.5$, $\mu=2$, $\Delta=5$, $t'=\bar{t}=10.2$, $\mu'=\bar{\mu}=\Delta'=\bar{\Delta}=0.2$, a choice of parameters equivalent to that of Figs. 1(c) and (h). Panels (a) and (c) show that the maximum spectral weight of the ring-Majoranas and of the Majorana bound state occurs at $\omega\sim\pm5$ [a.u.], which are the resonance energies displayed in Figs. 1(c)-(d). Panels (a)-(b) indicate that the spectral functions of ring-Majoranas is even with respect to $k$ for equal-sign Majoranas and with respect to $\omega$ for opposite-sign Majoranas in the Majorana bound state picture, whereas these functions do not have definite parity with respect to either $k$ or $\omega$ in the trivial bound state picture.}
\label{Fig2}
\end{center}
\end{figure}

\begin{figure}
\begin{center}
\includegraphics[width=1.0\columnwidth]{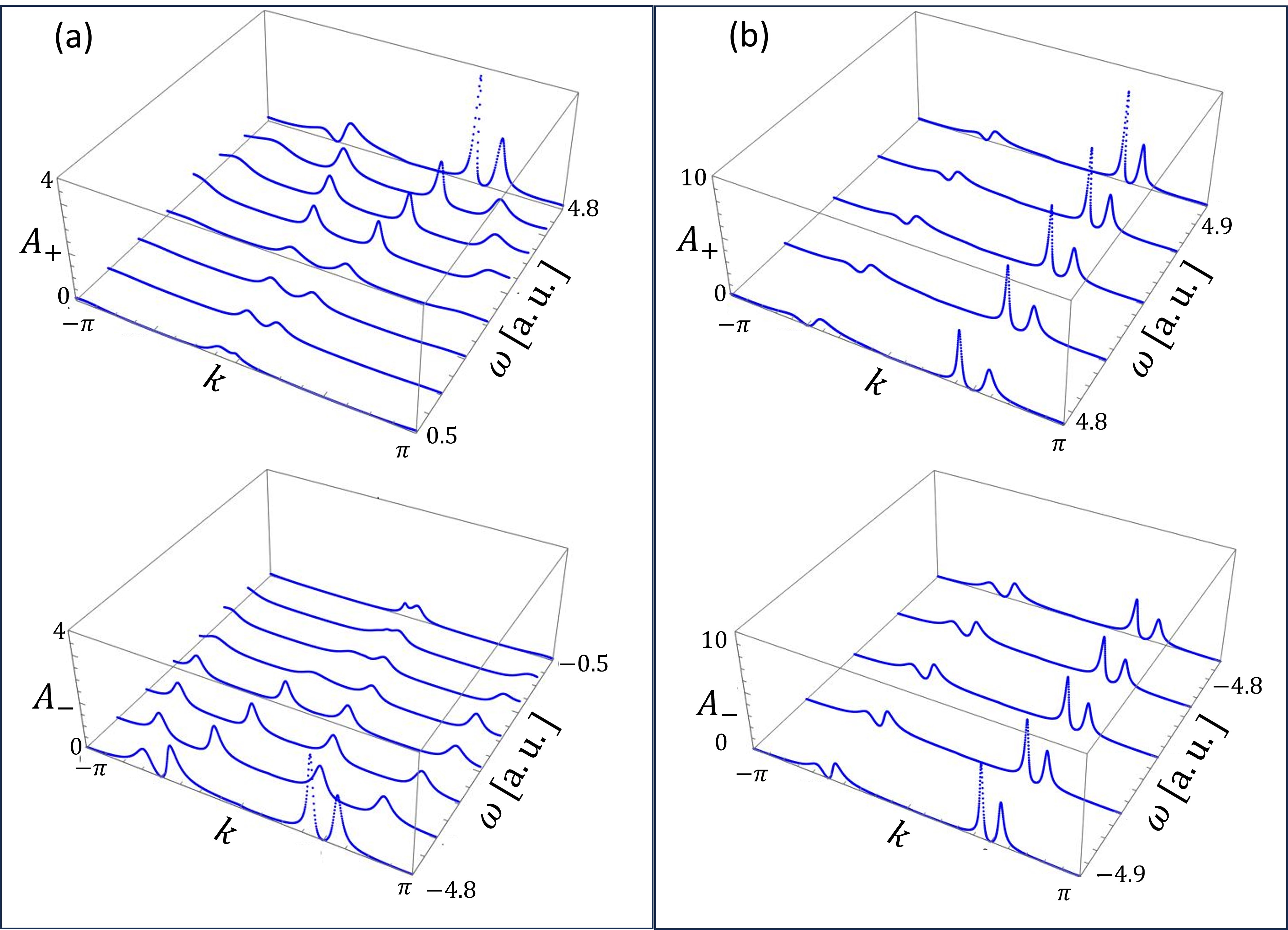}
\caption{Spectral functions $A_{+}$ and $A_{-}$ of the ring-Majoranas as a function of wave number $k$ (in units of lattice spacing) and energy $\omega$ in the trivial bound state picture, for adjacent intervals of $\omega$ in (a) and (b). As in Fig. 2(b), $t=1.5$, $\mu=2$, $\Delta=5$, $\bar{t}=10.2$, $\bar{\mu}=\bar{\Delta}=0.2$. The plots indicate that the lack of parity of $A_{+}$ and $A_{-}$ with respect to $k$ becomes more pronounced as $|\omega|$ increases, with a high concentration of spectral weight at the $k>0$ side at large $\omega$. Meanwhile, the lack of parity between $A_{+}$ and $A_{-}$ with respect to $\omega$ increases with decreasing $|\omega|$, with $A_{+}$ displaying significantly softened peaks when compared to $A_{-}$ near the edges of the Brillouin zone at small $\omega$.}
\label{Fig3}
\end{center}
\end{figure}

Figs. 4(a)-(b) show the expectation values of the occupation numbers of ring-Majoranas $\langle n_{\pm}(k) \rangle$ (c.f. Appendix A) in the MBS (panel (a)) and TBS (panel (b)) pictures. These results convey that while, in the MBS picture, $\langle n_{+}(k) \rangle=\langle n_{-}(k) \rangle$ are even functions of $k$ (the equality coming from $A_{-}(k,-\omega)=A_{+}(k,\omega)$ and the evenness from $A_{\pm}(-k,\omega)=A_{\pm}(k,\omega)$), in the TBS picture, $\langle n_{+}(k) \rangle\neq\langle n_{-}(k) \rangle$ do not have a definite parity (the inequality coming from $A_{-}(k,-\omega)\neq A_{+}(k,\omega)$ and the lack of parity from $A_{\pm}(-k,\omega)\neq A_{\pm}(k,\omega)$). Figs. 4(c)-(d), which show the expectation values of the occupation numbers of particles and holes $\langle n_{\text{p}/\text{h}}(k) \rangle$ (c.f. Appendix A) in the MBS (panel (c)) and TBS (panel (d)) pictures, indicate that the aforementioned parity properties of $\langle n_{\pm}(k) \rangle$ carry over to $\langle n_{\text{p}/\text{h}}(k) \rangle$ (which stems from the mathematical relation between these quantities). We thus see that the ring-TBS coupling induces time-reversal symmetry breaking ($\langle n_{\text{p}/\text{h}}(-k) \rangle\neq\langle n_{\text{p}/\text{h}}(k) \rangle$) and particle-hole symmetry breaking ($\langle n_{\text{p}}(k) \rangle\neq\langle n_{\text{h}}(k) \rangle$) in the occupation numbers of particles and holes, although their spectra possess those symmetries.
\begin{figure}
\begin{center}
\includegraphics[width=1.0\columnwidth]{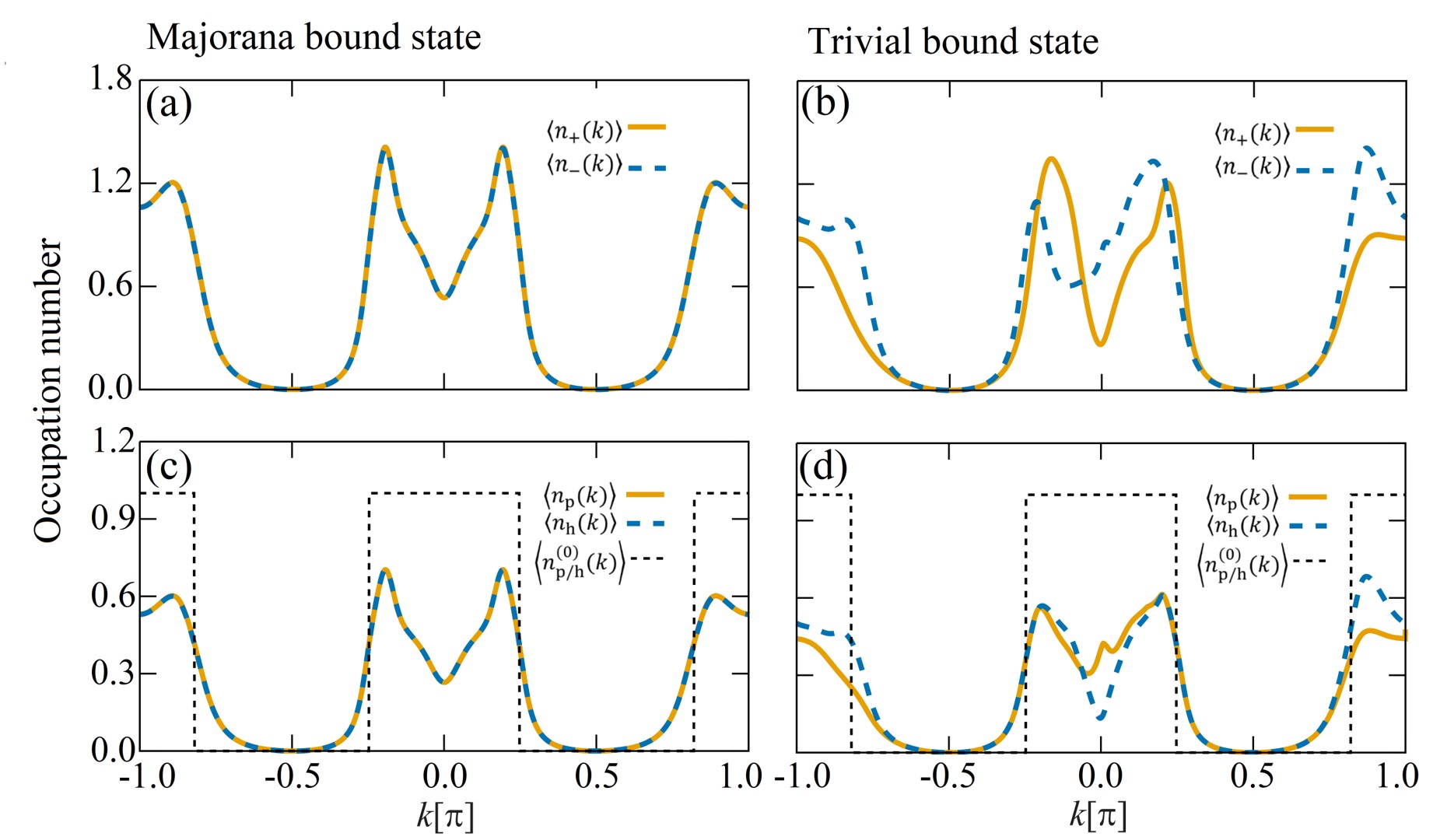}
\caption{(a) [(b)] The expectation values of the occupations numbers of ring-Majoranas $\langle n_{\pm}(k) \rangle$ are equal and even functions [unequal and do not have a definite parity] with respect to the wave number $k$ (in units of lattice spacing) in the Majorna bound state (MBS) [trivial bound state (TBS)] picture. (c) [(d)] The expectation values of the occupations numbers of particle and holes $\langle n_{\text{p}/\text{h}}(k) \rangle$ inherit the parity properties of $\langle n_{\pm}(k) \rangle$, thus implying time-reversal and particle-hole symmetry preservation [breaking] in the occupation of particle and holes in the MBS [TBS] picture. For reference, $\langle n_{\text{p}/\text{h}}^{(0)}(k) \rangle=\theta(E_{\text{F}}-E(k))$ is the zero-temperature and zero-coupling Fermi-Dirac distribution for particles and holes as a function of $k$, with $\pm E(k)=\pm2|f(k)|$ the particles and holes' bands at zero coupling and $\pm E_{\text{F}}$ their Fermi energies. As in Figs. 2(a)-(b), $t=1.5$, $\mu=2$, $\Delta=5$, $t'=\bar{t}=10.2$, $\mu'=\bar{\mu}=\Delta'=\bar{\Delta}=0.2$, and the Fermi energy was taken as 3.5 [a.u] when integrating the ring-Majoranas spectral functions in $\omega$ to get $\langle n_{\pm}(k) \rangle$.}
\label{Fig4}
\end{center}
\end{figure}

Figs. 5(a)-(b) displays the particle and hole currents $j_{\text{p}/\text{h}}$ in the ring (c.f. Appendix A) in the MBS (dashed orange and dashed blue) and TBS (orange and blue) pictures as a function of a ring-edge coupling parameter - $t'$ in the MBS picture and $\bar{t}$ in the TBS picture (panel (a)), and $\Delta'$ in the MBS picture and $\bar{\Delta}$ in the TBS picture (panel (b)). While in the MBS picture the preservation of time-reversal and particle-hole symmetries in $\langle n_{\text{p}/\text{h}}(k) \rangle$ results in the vanishing of $j_{\text{p}/\text{h}}$, the breaking of those symmetries wields nonzero and unbalanced $j_{\text{p}/\text{h}}$ in the TBS picture. These spontaneous currents are an interesting consequence of breaking the parities of the spectral functions due to nontrivial commutation relations between the edge-Majorana operators and the edge-Hamiltonian in the TBS picture.
\begin{figure}
\begin{center}
\includegraphics[width=\linewidth]{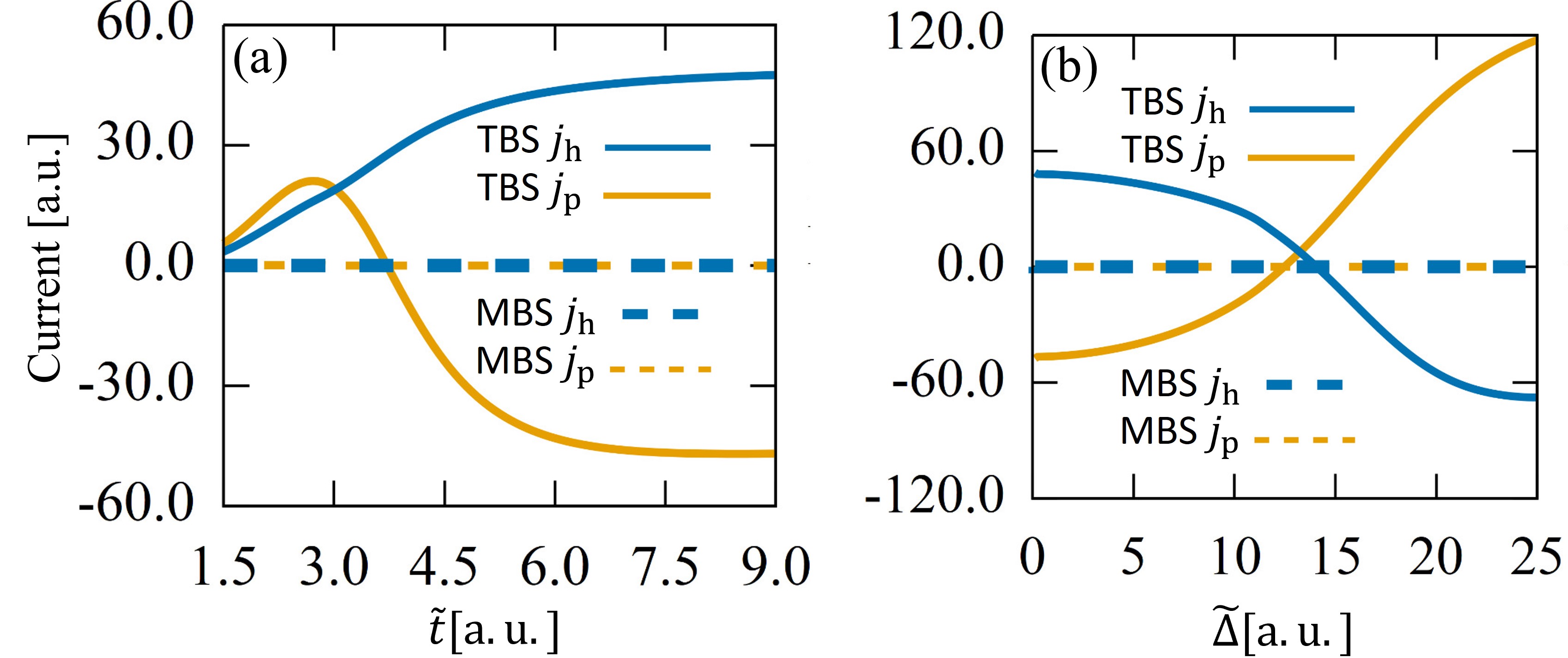}
\caption{(a) [(b)] Particle and hole currents $j_{\text{p}/\text{h}}$ as a function of ring-edge coupling parameter $\tilde{t}=t'$ [$\tilde{\Delta}=\Delta'$] vanish in the Majorana bound state (MBS) picture, while particle and hole currents $j_{\text{p}/\text{h}}$ as a function of ring-edge coupling parameter $\tilde{t}=\bar{t}$ [$\tilde{\Delta}=\bar{\Delta}$] are nonzero in the trivial bound state (TBS) picture. The fixed parameters are taken as in Fig. 4.}
\label{Fig4}
\end{center}
\end{figure}

\section{Conclusion}

We found that coupling a Kitaev ring to a Majorana bound state (MBS) or to a zero-energy trivial bound state (TBS) yields different band structures owing to the chiral symmetry of the uncoupled ring being broken by the coupling to a MBS, but partially preserved by the coupling to a TBS. In the MBS picture, the combination of energy resonance between the hybridize states and the maximum in the ring-Majoranas' and MBS spectral functions exactly at the resonance energy points to a high leakage probability between the resonating states. The same does not occur in the TBS picture where there is no energy resonance between the states with enhanced spectral weight.

The MBS and TBS scenarios are also strikingly different regarding the parities of ring-Majoranas spectral functions which, in the MBS picture, come out even with respect to wave number for like-Majoranas and with respect to energy for Majoranas of different species, whereas these functions do not possess definite parities in the TBS picture. As a consequence, time-reversal and particle-hole symmetries are broken in the occupation of particles and holes and currents are spontaneously generated in the TBS picture.

Varying the chemical potential in the Kitaev open chain across its critical value using a backgate, the emergence of a MBS in one of the ends of this chain will thus be signaled in the ancillary Kitaev ring. When the Kitaev chain is in the trivial phase and, hence, the Kitaev ring is coupled to a TBS, the ring's spectral functions and occupation numbers display non-definite parities and the ring supports a nonzero current. But when the Kitaev chain enters the topological phase, implying coupling of the Kitaev ring to a MBS, the spectral functions and occupation numbers acquire definite parities and the current in the ring drops to zero. The simple scheme of Fig. 1(a) thus serves as an unambiguous MBS-detector.

Combining two or more of the proposed scheme and exploiting the leakage probability between ring- and edge-Majorana states in the MBS picture could, conceivably, yield a platform for braiding and/or fusing these states, with relevant impacts for both Majorana detection and quantum computing.

\section*{Acknowledgements}

The authors acknowledge fruitful discussion with Igor Zutic and Alexandre Dodonov.
This work is supported by Fundacao de Apoio a Pesquisa do Distrito Federal (FAPDF), grant no. 00193-00001817/2023-43 and grant no. 00193-00002073/2023-84 and US ONR under awards MURI N000142212764, N000141712793.

\begin{appendix}

\section{Calculations for the Green's functions, spectral functions, occupation numbers, and currents}

To compute the ring-Majoranas Green's functions, $G_{\alpha}(k,\omega^{+})\,\equiv\,\ll\gamma_{\alpha,k}:\gamma_{\alpha,k}^{\dagger}\gg_{\omega^{+}}$, $\alpha=\pm$, we employ the equation of motion for the Green's function using Zubarev's formalism \cite{Zubarev1960},
\begin{equation}\label{EqMotion}
\nonumber 4\omega^{+}\ll\gamma_{\eta'}:\gamma_{\eta}^{\dagger}\gg_{\omega^{+}}=<\{\gamma_{\eta'},\gamma_{\eta}^{\dagger}\}>+\ll[\gamma_{\eta'},H]:\gamma_{\eta}^{\dagger}\gg_{\omega^{+}},
\end{equation}
with $\eta=\alpha,k$ encoding the ring-Majoranas' indexes and $H$ [$H=\bar{H}$] given in Eqs. (4)-(5) [Eqs. (8)-(9)] for the MBS [TBS] picture, along with the relations: $\gamma_{\alpha,k}^{\dagger}=-\gamma_{-\alpha,-k}$, $\{\gamma_{\alpha,k},\gamma_{\alpha',k'}\}=\{\gamma_{\alpha,k}^{\dagger},\gamma_{\alpha',k'}^{\dagger}\}=-2\delta_{-k,k'}\delta_{-\alpha,\alpha'}$, $\{\gamma_{\alpha,k}^{\dagger},\gamma_{\alpha',k'}\}=2\delta_{k,k'}\delta_{\alpha,\alpha'}$. The procedure yields $G_{\alpha}\,=\,G_{\alpha}^{(0)}\,+\,G_{\alpha}^{(0)^{2}}\,\Sigma$, where
\begin{equation}\label{Galpha0}
G_{\alpha}^{(0)}\,=\,G_{\alpha}^{(0)}(k,\omega^{+})\,=\,\frac{0.5}{\omega^{+}-\alpha|f(k)|},
\end{equation}
is the Green's function of the uncoupled ring and
\begin{equation}\label{Sigma}
\Sigma\,=\,\Sigma(k,\omega^{+})\,=\,\frac{0.5|h(k)|^{2}}{F(\omega^{+})},
\end{equation}
\begin{eqnarray}\label{Sigmaalpha}
\nonumber \Sigma\,&=&\,\Sigma_{\alpha}(k,\omega^{+})\,=\\
&=&\,\frac{0.5|\bar{v}(k)|^{2}\{C(\omega^{+})-\frac{\alpha}{|f(k)|}\text{Re}[f(k)D(\omega^{+})]\}}{|C(\omega^{+})|^{2}-|D(\omega^{+})|^{2}},
\end{eqnarray}
are the self-energies in the MBS and TBS pictures, respectively. In Eqs. (\ref{Sigma}) and (\ref{Sigmaalpha}),
\begin{equation}\label{h}
h(k)=2[\omega_{+}(k)+\omega_{-}(k)],
\end{equation}
\begin{equation}\label{F}
F(\omega^{+})=4\omega^{+}-\sum_{k',\alpha'}\frac{0.25|h(k')|^2}{\omega^{+}-\alpha'|f(k')|},
\end{equation}
\begin{equation}\label{C}
C(\omega^{+})=4\omega^{+}-\sum_{k',\alpha'}\frac{0.125|\bar{v}(k')|^2}{\omega^{+}-\alpha'|f(k')|},
\end{equation}
\begin{equation}\label{D}
D(\omega^{+})=i\bar{\varepsilon}_{\text{e}}-\sum_{k',\alpha'}\alpha'\frac{0.125|\bar{\bar{v}}(k')|^2}{\omega^{+}-\alpha'|f(k')|},
\end{equation}
where $|\bar{\bar{v}}(k)|^{2}\,=\,|\bar{v}(k)|^{2}f^{\ast}(k)/|f(k)|$. Plugging $\omega_{\pm}(k)=(-2t_{\pm}\cos k+\mu_{\pm})\,-\,i(2\Delta_{\pm}\sin k)$ in Eq. (\ref{h}) yields that $h(k)$ itself is of the form $h(k)=(-2t'\cos k+\mu')\,-\,i(2\Delta'\sin k)$, where $x'\,=\,2(x_{+}+x_{-})$, with $x\,=\,t,\mu,\Delta$.

With the same procedure used for the ring-Majoranas Green's functions, we obtain the MBS Green's function, $G_{\text{e}}(\omega^{+})\,\equiv\,\ll\gamma_{\text{e}}:\gamma_{\text{e}}^{\dagger}\gg_{\omega^{+}}$, in the MBS picture:
\begin{equation}\label{Ge}
G_{\text{e}}(\omega^{+})\,=\,\frac{2}{F(\omega^{+})}.
\end{equation}

Using Dyson equation and taking the limit $\eta\rightarrow0$ in $\omega^{+}=\omega+i\eta$, the ring-Majoranas Green's functions become, in both pictures, $G_{\alpha}\,=\,\text{Re}\,G_{\alpha}+i\,\text{Im}\,G_{\alpha}$ with
\begin{equation}\label{Re}
\text{Re}\,G_{\alpha}(k,\omega)\,=\,\frac{0.5[\omega-\alpha|f(k)|-0.5M(k,\omega)]}{[\omega-\alpha|f(k)|-0.5M(k,\omega)]^{2}+[0.5\Lambda(k,\omega)]^{2}},
\end{equation}
\begin{equation}\label{Im}
\text{Im}\,G_{\alpha}(k,\omega)\,=\,\frac{0.25\Lambda(k,\omega)}{[\omega-\alpha|f(k)|-0.5M(k,\omega)]^{2}+[0.5\Lambda(k,\omega)]^{2}},
\end{equation}
where $M[\Lambda]\,=\,\text{Lim}_{\eta\rightarrow0}\,\text{Re[Im]}\,\Sigma$ are given by
\begin{equation}\label{M}
M(k,\omega)\,=\,\frac{0.5|h(k)|^{2}F(\omega)}{F^{2}(\omega)+G^{2}(\omega)},
\end{equation}
\begin{equation}\label{Lambda}
\Lambda(k,\omega)\,=\,\frac{0.5|h(k)|^{2}G(\omega)}{F^{2}(\omega)+G^{2}(\omega)},
\end{equation}
\begin{equation}\label{Malpha}
M_{\alpha}(k,\omega)\,=\,\frac{0.5|\bar{v}(k)|^{2}[F_{\alpha}(k,\omega)F_{1}(\omega)+\bar{G}(\omega)F_{2}(\omega)]}{F_{1}^{2}(\omega)+F_{2}^{2}(\omega)},
\end{equation}
\begin{equation}\label{Lambdaalpha}
\Lambda_{\alpha}(k,\omega)\,=\,\frac{0.5|\bar{v}(k)|^{2}[\bar{G}(\omega)F_{1}(\omega)-F_{\alpha}(k,\omega)F_{2}(\omega)]}{F_{1}^{2}(\omega)+F_{2}^{2}(\omega)}
\end{equation}
in the MBS and TBS pictures, respectively. In Eqs. (\ref{Malpha})-(\ref{Lambdaalpha}),
\begin{equation}\label{Falpha}
F_{\alpha}(k,\omega)\,=\,C(\omega)-\frac{\alpha}{|f(k)|}\text{Re}\{f(k)[D(\omega)+i\bar{\bar{G}}(\omega)]\},
\end{equation}
\begin{equation}\label{F2}
F_{1}(\omega)\,=\,C^{2}(\omega)-\bar{G}^{2}(\omega)-|D(\omega)+i\bar{\bar{G}}(\omega)|^{2},
\end{equation}
\begin{equation}\label{F3}
F_{2}(\omega)\,=\,2C(\omega)\bar{G}(\omega).
\end{equation}

Function $F$, featured in Eqs. (\ref{M})-(\ref{Lambda}), and functions $C$ and $D$, in Eqs. (\ref{Falpha})-(\ref{F3}), are given in the Appendix B. It is found that $F$ and $C$ are odd functions of $\omega$, while $D$ is even. Function $G$, featured in Eqs. (\ref{M})-(\ref{Lambda}), and functions $\bar{G}$ and $\bar{\bar{G}}$, in Eqs. (\ref{Falpha})-(\ref{F3}), are given in Appendix C. Whereas $G$ and $\bar{G}$ are even, $\bar{\bar{G}}$ is odd.

Eqs. (\ref{M})-(\ref{Lambda}) and the aforementioned parities of $F$ and $G$ yields that $M$ and $\Lambda$ are both even in $k$, while $M$ is odd and $\Lambda$ is even in $\omega$ in the MBS picture. Combining these facts and Eq. (\ref{Im}) yields that $\text{Im}\,G_{\alpha}(-k,\omega)=\text{Im}\,G_{\alpha}(k,\omega)$ and $\text{Im}\,G_{-\alpha}(k,-\omega)=\text{Im}\,G_{\alpha}(k,\omega)$ in the MBS picture. Meanwhile, Eqs. (\ref{Falpha})-(\ref{F3}) and the parities of $C$, $D$, $\bar{G}$, and $\bar{\bar{G}}$ yield that while $F_{2}$ is odd, $F_{1}$ and $F_{\alpha}$ do not have a definite parity, the latter neither in $\omega$ nor in $k$. It follows that functions $M_{\alpha}$ and $\Lambda_{\alpha}$ in Eqs. (\ref{Malpha})-(\ref{Lambdaalpha}) and, hence, $\text{Im}\,G_{\alpha}$ in Eq. (\ref{Im}) do not have definite parity relations in $\omega$ or in $k$ in the TBS picture.

The $\eta\rightarrow0$ MBS Green's function in the MBS picture is $G_{\text{e}}\,=\,\text{Re}\,G_{\text{e}}+i\,\text{Im}\,G_{\text{e}}$ with
\begin{equation}\label{Ree}
\text{Re}\,G_{\text{e}}(\omega)\,=\,\frac{2F(\omega)}{F^{2}(\omega)+G^{2}(\omega)},
\end{equation}

\begin{equation}\label{Ime}
\text{Im}\,G_{\text{e}}(\omega)\,=\,\frac{2G(\omega)}{F^{2}(\omega)+G^{2}(\omega)}.
\end{equation}

The particles ($\text{p}$) and holes ($\text{h}$) Green's functions, $G_{\text{p/h}}$, read in both pictures
\begin{equation}\label{Gph}
G_{\text{p}/\text{h}}(k,\omega_{\text{p}/\text{h}})\,=\,|g_{+}(k)|^{2}\,G_{\pm}(k,\omega_{\pm})\,+\,|g_{-}(k)|^{2}\,G_{\mp}(k,\omega_{\mp}),
\end{equation}
where the upper/lower sign holds for $\text{p}/\text{h}$ and $|g_{\pm}(k)|^{2}=(1\pm\text{Im}\,f(k)/|f(k)|)/4$. The energy $\omega_{\text{p}/\text{h}}$ of the particle/hole is mapped from the energies $\omega_{\pm}$ of the ring-Majoranas: $\omega_{\text{p}/\text{h}}=c_{\text{p}/\text{h}}^{+}\omega_{+}+c_{\text{p}/\text{h}}^{-}\omega_{-}$, with the constraint that $\omega_{+}>0\Leftrightarrow\omega_{-}<0$ and $\omega_{+}<0\Leftrightarrow\omega_{-}>0$, and with coefficients $c_{\text{p}/\text{h}}^{\pm}$ such that the positive and negative ring-Majoranas energy sectors are mapped onto the correct particle and hole energy sectors, as depicted in Fig. 6 for the particle case.
\begin{figure}
\begin{center}
\includegraphics[width=1.0\columnwidth]{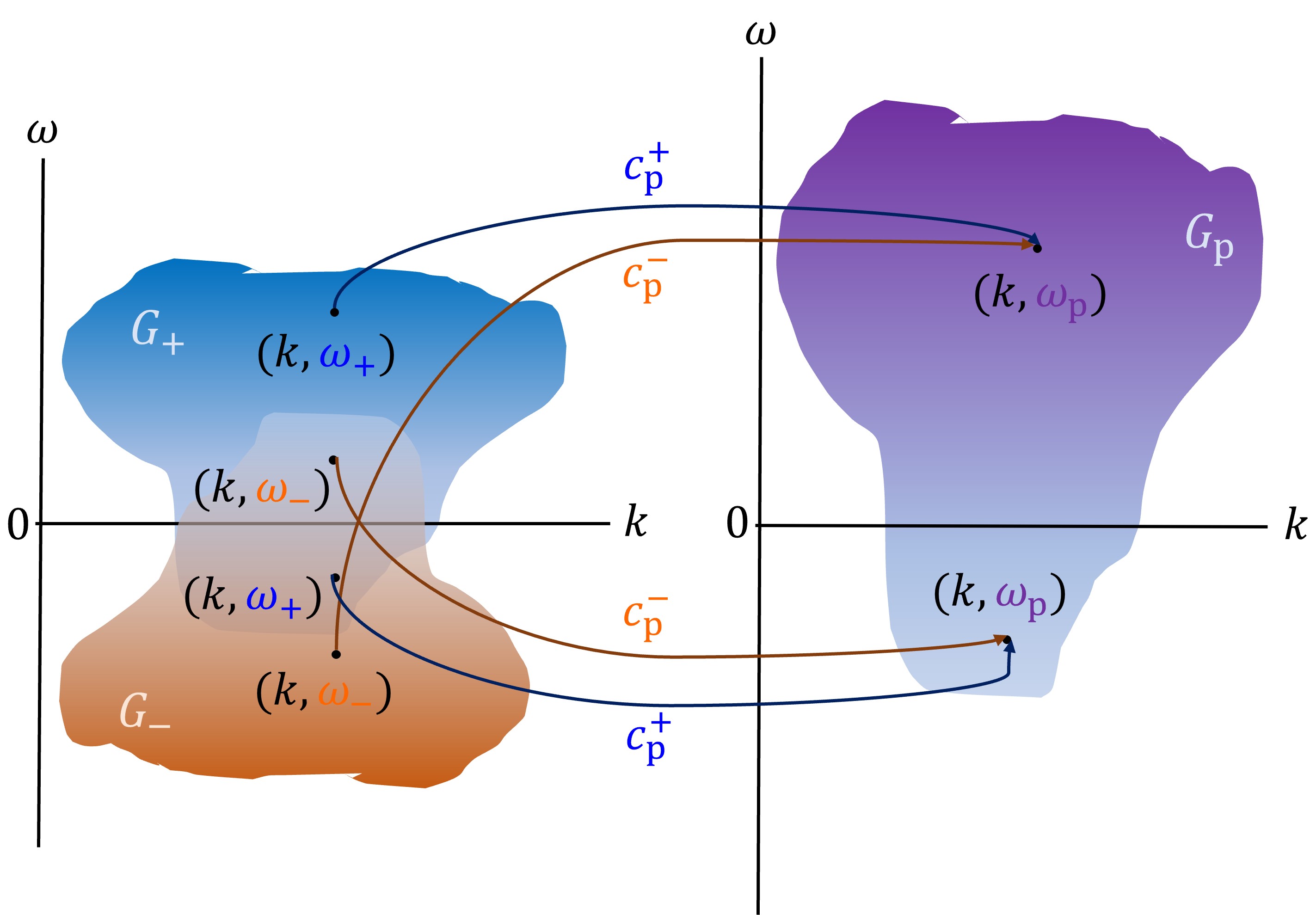}
\caption{Mapping between the domains of the particles Green's function $G_{\text{p}}(k,\omega_{\text{p}})$ and the domains of the ring-Majoranas' Green's functions $,G_{\pm}(k,\omega_{\pm})$.}
\label{Fig6}
\end{center}
\end{figure}

Eq. (\ref{Gph}) implies that the spectral functions, $A_{\eta}=-\text{Im}\,G_{\eta}$, follow the same relation:
\begin{equation}\label{Aph}
A_{\text{p}/\text{h}}(k,\omega_{\text{p}/\text{h}})\,=\,|g_{+}(k)|^{2}\,A_{\pm}(k,\omega_{\pm})\,+\,|g_{-}(k)|^{2}\,A_{\mp}(k,\omega_{\mp}).
\end{equation}

The expectation value of the occupation number of particles/holes, $\langle n_{\text{p}/\text{h}} \rangle$, is given by
\begin{equation}\label{occupation}
\langle n_{\text{p}/\text{h}}(k) \rangle=\int_{-\infty}^{+\infty}A_{\text{p}/\text{h}}(k,\omega_{\text{p}/\text{h}})\,n_{\text{p}/\text{h}}^{\text{F}}(\omega_{\text{p}/\text{h}})\,d\omega_{\text{p}/\text{h}},
\end{equation}
where $n_{\text{p}/\text{h}}^{\text{F}}$ is the particle/hole Fermi-Dirac distribution which, at zero temperature, is given by
\begin{equation}\label{occupationzeroT}
n_{\text{p}/\text{h}}^{\text{F}}(\omega_{\text{p}/\text{h}})=\theta(\pm E_{\text{p}/\text{h}}^{\text{F}}\mp\omega_{\text{p}/\text{h}}),
\end{equation}
with upper/lower sign for $\text{p}/\text{h}$, $\theta$ the Heaviside step function, and $E_{\text{p}/\text{h}}^{\text{F}}$ the Fermi energy at the particle/hole band ($E_{\text{h}}^{\text{F}}=-E_{\text{p}}^{\text{F}}=-E^{\text{F}}$ for particle-hole symmetric bands).

At zero temperature and zero edge-ring coupling, in which case $A_{\text{p}/\text{h}}(k,\omega_{\text{p}/\text{h}})=A_{\text{p}/\text{h}}^{(0)}(k,\omega_{\text{p}/\text{h}})=\delta(\omega_{\text{p}/\text{h}}-E_{\text{p}/\text{h}}(k))$ and $E_{\text{h}}(k)=-E_{\text{p}}(k)=-E(k)$, Eqs. (\ref{occupation})-(\ref{occupationzeroT}) yield
\begin{equation}\label{occupation0}
\langle n_{\text{p}/\text{h}}^{(0)}(k) \rangle=\theta(E_{\text{F}}-E(k)).
\end{equation}

At zero temperature and nonzero edge-ring coupling, Eqs. (\ref{occupation})-(\ref{occupationzeroT}) give
\begin{equation}\label{occupation2}
\langle n_{\text{p}/\text{h}}(k) \rangle=\int_{a_{\text{p}/\text{h}}}^{b_{\text{p}/\text{h}}}A_{\text{p}/\text{h}}(k,\omega_{\text{p}/\text{h}})\,d\omega_{\text{p}/\text{h}},
\end{equation}
where $a_{\text{p}}=-\infty$, $b_{\text{p}}=E_{\text{p}}^{\text{F}}$, $a_{\text{h}}=E_{\text{h}}^{\text{F}}$, $b_{\text{h}}=\infty$.

Inserting Eq. (\ref{Aph}) into Eq. (\ref{occupation2}) (recalling the mapping between the integration variable $\omega_{\text{p}/\text{h}}$ in Eq. (\ref{occupation2}) and $\omega_{\pm}$ in Eq. (\ref{Aph}) to change accordingly the integration variables and the limits of the integrals), we get
\begin{eqnarray}\label{occupation3}
\nonumber \langle n_{\text{p}/\text{h}}(k) \rangle\,&=&\,|g_{+}(k)|^{2}\int_{a_{\pm}}^{b_{\pm}}A_{\pm}(k,\omega_{\pm})\,d\omega_{\pm}+\\
&+&|g_{-}(k)|^{2}\int_{a_{\mp}}^{b_{\mp}}A_{\mp}(k,\omega_{\mp})\,d\omega_{\mp},
\end{eqnarray}
where $a_{+}=-\infty$, $b_{+}=E_{+}^{\text{F}}$, $a_{-}=E_{-}^{\text{F}}$, $b_{-}=\infty$.

Let us define
\begin{equation}\label{occupationpm}
\langle n_{\pm}(k) \rangle\,=\,\int_{a_{\pm}}^{b_{\pm}}A_{\pm}(k,\omega)\,d\omega,
\end{equation}
where we have removed from $\omega$ the now unnecessary lower indexes (since these energies are being integrated out), and call them ``expectation values of the occupation numbers of ring-Majoranas" (mind that these are but given names since the ring-Majoranas do not follow regular fermionic statistics and, hence, they do not have a step-function zero-temperature occupation equivalent to Eq. (\ref{occupationzeroT}), with the result that an equivalent of Eq. (\ref{occupation}) for the ring-Majoranas would not reduce to an equivalent of Eq. (\ref{occupation2})). With that Eq. (\ref{occupation3}) writes
\begin{equation}\label{occupation4}
\langle n_{\text{p}/\text{h}}(k) \rangle\,=\,|g_{+}(k)|^{2}\langle n_{\pm}(k) \rangle+|g_{-}(k)|^{2}\langle n_{\mp}(k) \rangle,
\end{equation}

When the ring-Majoranas' spectral functions obey the parity relation $A_{\pm}(-k,\omega)=A_{\pm}(k,\omega)$, Eq. (\ref{occupationpm}) renders $\langle n_{\pm}(k) \rangle$ an even function of $k$ which, by virtue of Eq. (\ref{occupation4}) and the fact that $|g_{\pm}(k)|^{2}$ are even, implies $\langle n_{\text{p}/\text{h}}(k) \rangle$ also even. Now, when $A_{-}(k,-\omega)=A_{+}(k,\omega)$, Eq. (\ref{occupationpm}) yields $\langle n_{+}(k) \rangle=\langle n_{-}(k) \rangle=\langle n(k) \rangle$ and Eq. (\ref{occupation4}) becomes simply
\begin{equation}\label{occupation5}
\langle n_{\text{p}/\text{h}}(k) \rangle\,=\,(|g_{+}(k)|^{2}+|g_{-}(k)|^{2})\langle n(k) \rangle=0.5\langle n(k) \rangle,
\end{equation}
i.e., the expectation values of the occupation numbers of particles and holes are equal and are given by the integral of the spectral function of either species of ring-Majoranas. But if $A_{-}(k,-\omega)\neq A_{+}(k,\omega)$, then $\langle n_{+}(k) \rangle\neq\langle n_{-}(k) \rangle$ and $\langle n_{\text{p}/\text{h}}(k) \rangle$, which are the combination of these integrals weighted by $|g_{\pm}(k)|^{2}$, differ by $0.5\text{Im}\,f(k)/|f(k)|[\langle n_{+}(k) \rangle-\langle n_{-}(k) \rangle]$.

The currents generated by particles/holes, $j_{\text{p}/\text{h}}$, is
\begin{equation}\label{currentph}
j_{\text{p}/\text{h}}\,=\,\mp\frac{e\hbar}{m}\sum_{k}k\,\langle n_{\text{p}/\text{h}}(k) \rangle,
\end{equation}
where the upper/lower sign holds for $\text{p}/\text{h}$, $e$ is the modulus of the particle/hole charge, $m$ its mass. (In a superconductor, particles and holes are the excitations upon the ground state of Cooper pairs whose energy lies inside the gap between the particle and hole bands. Therefore, these particles and holes have the same mass, unlike, e.g., in semiconductors where electrons and holes have different effective masses owing to their different roles in the crystalline structure.) If $\langle n_{\text{p}/\text{h}}(k) \rangle$ is an even function of $k$, the sum in Eq. (\ref{currentph}) vanishes. Else, an imbalance in the number os particles/holes with positive and with negative momenta wields a nonzero current. If $\langle n_{\text{p}}(k) \rangle\neq\langle n_{\text{h}}(k) \rangle$ (and are not even), Eq. (\ref{currentph}) wields $\langle j_{\text{p}}\neq j_{\text{h}} \rangle$.

\section{Analytical computation of functions entering the Green's functions using residues}

Functions $F(\omega)$, $C(\omega)$, and $D(\omega)$ are defined by Eqs. (\ref{F})-(\ref{D}) through the replacement $\omega^{+}\rightarrow\omega\in\Re$:
\begin{equation}\label{Fofomega}
F(\omega)=4\omega-\sum_{k,\alpha}\frac{0.25|h(k)|^{2}}{\omega-\alpha|f(k)|},
\end{equation}
\begin{equation}\label{Cofomega}
C(\omega)=4\omega-\sum_{k,\alpha}\frac{0.125|\bar{v}(k)|^{2}}{\omega-\alpha|f(k)|},
\end{equation}
\begin{equation}\label{Dofomega}
D(\omega)=i\bar{\varepsilon}_{\text{e}}-\sum_{k,\alpha}\alpha\frac{0.125|\bar{\bar{v}}(k)|^{2}}{\omega-\alpha|f(k)|},
\end{equation}
where here we have removed the tilde from the summed variables since there is no ambiguity.

Let
\begin{equation}\label{Salphaofomega}
S_{\alpha}(\omega)=\sum_{k}\frac{g(k)}{\omega-\alpha|f(k)|}=\int_{-\pi}^{\pi}\frac{dk}{2\pi}\,\frac{g(k)}{\omega-\alpha|f(k)|},
\end{equation}
where $g(k)=0.25|h(k)|^{2},0.125|\bar{v}(k)|^{2},0.125|\bar{\bar{v}}(k)|^{2}$. The last equality in Eq. (\ref{Salphaofomega}) holds in the thermodynamic limit of a crystal with $N\rightarrow\infty$ sites for which the separation between $k$'s vanishes as $N^{-1}$. This limit is necessary since, at the energy $\omega=\bar{\omega}$ and the momenta $k=\bar{k}\in[-\pi,\pi]$ (there might be more than one $\bar{k}$) for which $\bar{\omega}=\alpha|f(\bar{k})|$, the terms in the sum of Eq. (\ref{Salphaofomega}) diverge and, hence, $S_{\alpha}(\omega)\rightarrow\pm\infty$ when $\omega\rightarrow\bar{\omega}^{\pm}$. However, in the (continuous) thermodynamic limit the $\pm\infty$ cancel exactly, making $S(\omega)$ finite $\forall\,\omega$, as we shall see.

The integral over the real axis in Eq. (\ref{Salphaofomega}) becomes an integral on the complex plane through the parametrization $z=e^{ik}$:
\begin{equation}
\nonumber S_{\alpha}(\omega)=\int_{C}\frac{dz}{2\pi i}\,\frac{1}{z}\,\frac{g(z)}{\omega-\alpha|f(z)|}=\int_{C}\frac{dz}{2\pi i}\,f_{\alpha}(z),
\end{equation}
\begin{equation}\label{Salphaofomegathermo}
S_{\alpha}(\omega)=\sum_{z_{i}\,\text{in}\,C}\text{Res}[f_{\alpha},z_{i}],
\end{equation}
where $C$ is the unit circle traced out by $z=e^{ik}$ on the complex plane in the counterclockwise direction as $k$ swipes the real axis from $-\pi$ to $\pi$, and $f_{\alpha}(z)=g(z)/[\omega z-\alpha z|f(z)|]$. The Residue Theorem then yields Eq. (\ref{Salphaofomegathermo}), where $\text{Res}[f_{\alpha},z_{i}]$ are the residues of $f_{\alpha}(z)$ at its poles $z_{i}$ inside $C$.

The general prescription for the residue has that
\begin{equation}\label{Residue}
\text{Res}[f_{\alpha},z_{i}]=\frac{1}{(n-1)!}\lim_{z\rightarrow z_{i}}d_{z}^{n-1}[(z-z_{i})^{n}f_{\alpha}(z)],
\end{equation}
where $n$ is the multiplicity of $z_{i}$,

Let us start with functions $g(k)=0.25|h(k)|^{2},0.125|\bar{v}(k)|^{2}$, with $h(k)=(-2t'\cos k+\mu')\,-\,i(2\Delta'\sin k)$ and $\bar{v}(k)=(-2\bar{t}\cos k+\bar{\mu})\,-\,i(2\bar{\Delta}\sin k)$, which can both be generically written as
\begin{equation}\label{gofk}
g(k)=\frac{\tilde{d}}{4}[(-2\tilde{t}\cos k+\tilde{\mu})^{2}+(2\tilde{\Delta}\sin k)^{2}],
\end{equation}
where $\tilde{d}=1\,[1/2]$ and $\tilde{x}=x'\,[\bar{x}]$, $x=t,\mu,\Delta$, when using $h(k)\,[\bar{v}(k)]$.  Meanwhile, $\alpha|f(k)|$, with $f(k)=0.5\,[\,2\Delta\sin k\,+\,i(-2t\cos k+\mu)\,]$, reads off
\begin{equation}\label{galphaofk}
\alpha|f(k)|=\alpha\{\frac{1}{4}[(-2t\cos k+\mu)^{2}+(2\Delta\sin k)^{2}]\}^{1/2}.
\end{equation}

Using $\sin^{2}k=1-\cos^{2}k$ and that $2\cos k=z+z^{-1}$, we arrive at
\begin{equation}\label{gofz}
g(z)=\frac{\tilde{d}}{z^{2}}[\tilde{a}z^{4}+\tilde{b}z^{3}+(2\tilde{a}+\tilde{c})z^{2}+\tilde{b}z+\tilde{a}]=\frac{\tilde{N}(z)}{z^{2}},
\end{equation}
\begin{equation}\label{galphaofz}
\alpha|f(z)|=\alpha\frac{1}{z}[az^{4}+bz^{3}+(2a+c)z^{2}+bz+a]^{1/2}=\alpha\frac{N^{1/2}(z)}{z},
\end{equation}
where $\tilde{a}=(\tilde{t}^{2}-\tilde{\Delta}^{2})/4$, $\tilde{b}=-\tilde{t}\tilde{\mu}/2$, $\tilde{c}=\tilde{\Delta}^{2}+\tilde{\mu}^{2}/4$, likewise for $a$, $b$, $c$,
\begin{equation}\label{Nofz}
\tilde{N}(z)=\tilde{d}[\tilde{a}z^{4}+\tilde{b}z^{3}+(2\tilde{a}+\tilde{c})z^{2}+\tilde{b}z+\tilde{a}],
\end{equation}
and $N(z)$ given by Eq. (\ref{Nofz}) with removed tilde with $d=1$.

It follows that
\begin{equation}\label{falphaofz}
f_{\alpha}(z)=\frac{g(z)}{\omega z-\alpha z|f(z)|}=\frac{\tilde{N}(z)}{z^{2}D_{\alpha}(z)},
\end{equation}
where
\begin{equation}\label{Dalphaofz}
D_{\alpha}(z)=\omega z-\alpha N^{1/2}(z).
\end{equation}

Eq. (\ref{falphaofz}) and the fact that $\tilde{N}(z)$ is a polynomial in $z$, implies that the poles of $f_{\alpha}(z)$ are the zeros of $z^{2}D_{\alpha}(z)$. Hence, $z_{0}=0$ is a pole of multiplicity 2 and the other poles are obtained by solving $D_{\alpha}(z)=0$. Using Eq. (\ref{Dalphaofz}) and $z=e^{ik}$, $D_{\alpha}(z)=0$ implies that $\omega e^{ik}=\alpha Re^{i\theta}$ for some real numbers $R>0$ and $\theta$, which constraints $\omega>0$ to $\alpha=+$ and $\omega<0$ to $\alpha=-$. Else, $D_{\alpha}(z)\neq0$ and $f_{\alpha}(z)$ has no poles other than $z_{0}=0$. Now, using Eq. (\ref{Nofz}) for $N(z)$ and Eq. (\ref{Dalphaofz}), $D_{\alpha}(z)=0$ also implies that
\begin{equation}\label{poles}
az^{4}+bz^{3}+(2a+c-\omega^{2})z^{2}+bz+a=0.
\end{equation}

Eq. (\ref{poles}) implies that the poles $z_{i}\neq0$, $i=1,..,4$, for $t\neq\pm\Delta$, and that $z_{i}=z_{i}(w)$ are even functions of $\omega$. Solving Eq. (\ref{poles}) with Mathematica, we find the poles and the conditions for which they are inside the unit circle $C$:
\begin{eqnarray}\label{conds}
\nonumber |z_{1}|&<&1\qquad\text{for}\,\,|\omega|<\omega_{1},\\
|z_{3}|&<&1\qquad\text{for}\,\,|\omega|>\omega_{3},\\
\nonumber |z_{2}|&<&1\qquad\text{for}\,\,|\omega|<\omega_{2}\,\,\text{and}\,\,|\omega|>\omega_{3},\\
\nonumber |z_{4}|&<&1\qquad\text{for}\,\,|\omega|\in\{\varnothing\},
\end{eqnarray}
where $\omega_{1,2}=t\mp\mu/2$ (with $-$ for $\omega_{1}$ and $+$ for $\omega_{2}$), and $\omega_{3}=\Delta\{1+\mu^2/[4(\Delta^2-t^2)]\}^{1/2}$. Since we are considering $t>\mu/2$ and $\mu>0$ (topological condition), $\omega_{1,2}>0$ and $\omega_{1}<\omega_{2}$. Additionally, taking $\Delta>t$, $\omega_{2}<\omega_{3}$. The absolute values of the poles as a function of $\omega$ are displayed in Fig. 7.
\begin{figure}
\begin{center}
\includegraphics[width=1.0\columnwidth]{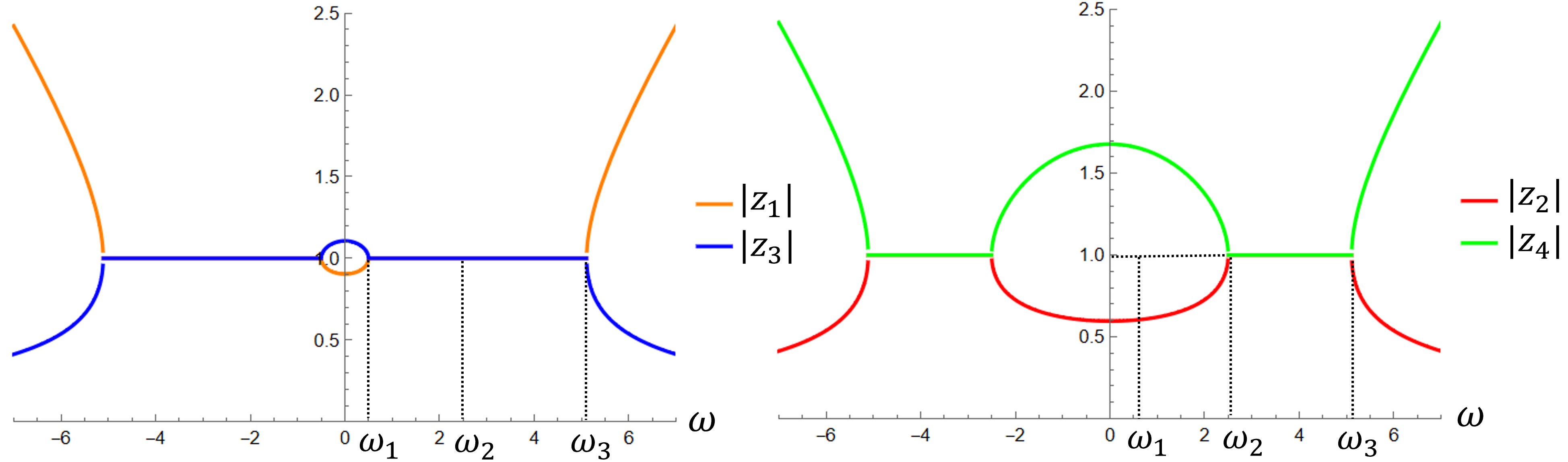}
\caption{Absolute values $|z_{i}|$ of the poles of $f_{\alpha}(z)$ as a function of $\omega$, $i=1,...,4$. The extremities of the intervals in $\omega$ are $\omega_{1,2}=t\mp\mu/2$ (with $-$ for $\omega_{1}$ and $+$ for $\omega_{2}$), and $\omega_{3}=\Delta\{1+\mu^2/[4(\Delta^2-t^2)]\}^{1/2}$. In this plot, $t=1.5$, $\mu=2$, $\Delta=5$.}
\label{Fig7}
\end{center}
\end{figure}

Also, we find that $z_{i}$, $i=1,...,4$, have multiplicity 1 $\forall\,\omega$, except when $|\omega|=\omega_{1},\omega_{2}$, in which case the poles $z_{1}=z_{3}=1$ and $z_{2}=z_{4}=-1$ have multiplicity 2.

Inputting into Eq. (\ref{Salphaofomegathermo}) the conditions of Eqs. (\ref{conds}) and the findings for the multiplicities yields
\begin{eqnarray}\label{Salphaofomegathermo2}
S_{\alpha}(\omega)&=&\text{Res}[f_{\alpha},z_{0}]+\\
\nonumber &+&\begin{cases}
                  \text{Res}[f_{\alpha},z_{1}]+\text{Res}[f_{\alpha},z_{2}], & \mbox{for } |\omega|<\omega_{1} \\
                  \text{Res}[f_{\alpha},z_{2}], & \mbox{for } \omega_{1}<|\omega|<\omega_{2} \\
                  \text{Res}[f_{\alpha},z_{2}]+\text{Res}[f_{\alpha},z_{3}], & \mbox{for } |\omega|>\omega_{3} \\
                  0, & \mbox{otherwise},
                \end{cases}
\end{eqnarray}
with $z_{0}$ of multiplicity 2 and $z_{1},\,z_{2},\,z_{3}$ of multiplicity 1.

We pass now to the evaluation of the residues. Employing Eq. (\ref{Residue}) with $n=2$ we get
\begin{equation}
\nonumber \text{Res}[f_{\alpha},z_{0}]=\lim_{z\rightarrow 0}d_{z}[z^{2}f_{\alpha}(z)],
\end{equation}
which, inserting Eq. (\ref{falphaofz}), yields
\begin{equation}\label{Residue0}
\text{Res}[f_{\alpha},z_{0}]=\lim_{z\rightarrow 0}\left[\frac{\tilde{N}'(z)}{D_{\alpha}(z)}-\frac{\tilde{N}(z)}{D_{\alpha}^{2}(z)}D'_{\alpha}(z)\right].
\end{equation}

Using Eqs. (\ref{Nofz}) and (\ref{Dalphaofz}) to obtain
\begin{equation}\label{dNtilde}
\tilde{N}'(z)=\tilde{d}[4\tilde{a}z^{3}+3\tilde{b}z^{2}+2(2\tilde{a}+\tilde{c})z+\tilde{b}],
\end{equation}
\begin{equation}\label{dDalpha}
D'_{\alpha}(z)=\omega-\frac{\alpha}{2N^{1/2}(z)}[4az^{3}+3bz^{2}+2(2a+c)z+b],
\end{equation}
and that $\lim_{z\rightarrow0}\tilde{N}(z)=\tilde{d}\tilde{a}$, $\lim_{z\rightarrow0}D_{\alpha}(z)=-\alpha\sqrt{a}$, $\lim_{z\rightarrow0}\tilde{N}'(z)=\tilde{d}\tilde{b}$, $\lim_{z\rightarrow0}D'_{\alpha}(z)=\omega-\alpha b/(2\sqrt{a})$, Eq. (\ref{Residue0}) becomes
\begin{equation}\label{Residue02}
\text{Res}[f_{\alpha},z_{0}]=-\frac{\tilde{d}\tilde{a}}{a}\omega+\alpha\frac{\tilde{d}}{\sqrt{a}}\left(\frac{\tilde{a}b}{2a}-\tilde{b}\right).
\end{equation}

Eq. (\ref{Residue}) with $n=1$ gives
\begin{equation}
\nonumber \text{Res}[f_{\alpha},z_{i}]=\lim_{z\rightarrow z_{i}}(z-z_{i})f_{\alpha}(z),
\end{equation}
which, inserting Eq. (\ref{falphaofz}) and employing the L'Hopital rule, yields
\begin{equation}\label{Residuei2}
\text{Res}[f_{\alpha},z_{i}]=\frac{\tilde{N}(z_{i})}{d_{z}[z^{2}D_{\alpha}(z)]_{z_{i}}}
\end{equation}
if $d_{z}[z^{2}D_{\alpha}(z)]_{z_{i}}\neq0$.

Now
\begin{equation}
\nonumber d_{z}[z^{2}D_{\alpha}(z)]_{z_{i}}=2z_{i}D_{\alpha}(z_{i})+z_{i}^{2}D'_{\alpha}(z_{i})=
\end{equation}
\begin{equation}
\nonumber =z_{i}^{2}[\omega-\frac{\alpha}{2N^{1/2}(z_{i})}(4az_{i}^{3}+3bz_{i}^{2}+2(2a+c)z_{i}+b)]=
\end{equation}
\begin{eqnarray}
\nonumber &=&z_{i}[\alpha N^{1/2}(z_{i})-\\
\nonumber &-&\frac{\alpha}{2N^{1/2}(z_{i})}(4az_{i}^{4}+3bz_{i}^{3}+2(2a+c)z_{i}^{2}+bz_{i})]=
\end{eqnarray}
\begin{equation}
\nonumber =\frac{\alpha z_{i}}{N^{1/2}(z_{i})}[N(z_{i})-(2az_{i}^{4}+\frac{3}{2}bz_{i}^{3}+(2a+c)z_{i}^{2}+\frac{b}{2}z_{i})]=
\end{equation}
\begin{equation}
\nonumber =\frac{\alpha z_{i}}{N^{1/2}(z_{i})}(-az_{i}^{4}-\frac{b}{2}z_{i}^{3}+\frac{b}{2}z_{i}+a),
\end{equation}
\begin{equation}\label{derivative}
d_{z}[z^{2}D_{\alpha}(z)]_{z_{i}}=\frac{\alpha z_{i}L(z_{i})}{2N^{1/2}(z_{i})},
\end{equation}
where
\begin{equation}\label{Lofz}
L(z)=-2az^{4}-bz^{3}+bz+2a.
\end{equation}

Equating Eq. (\ref{derivative}) to zero, combined with Eq. (\ref{poles}) with $z=z_{i}$, yields $z_{1}=z_{3}=1$ and $z_{2}=z_{4}=-1$ for $|\omega|=\omega_{1},\omega_{2}$, which are the poles of multiplicity 2 not contributing to Eq. (\ref{Salphaofomegathermo2}).

So plugging Eq. (\ref{derivative}) back into Eq. (\ref{Residuei2}) and incorporating the constraint that $\alpha=+[-]$ implies $\omega>0[<0]$ wields
\begin{equation}\label{Residuei3}
\text{Res}[f_{\alpha},z_{i}]=\alpha\,R(z_{i})\,\delta_{\alpha,\text{sign}(\omega)},
\end{equation}
where
\begin{equation}
\nonumber R(z_{i})=\frac{2\tilde{N}(z_{i})N^{1/2}(z_{i})}{z_{i}L(z_{i})}
\end{equation}
which, using Eqs. (\ref{gofz})-(\ref{galphaofz}), reads
\begin{equation}\label{Rzi}
R(z_{i})=\frac{2z_{i}^{2}g(z_{i})|f(z_{i})|}{L(z_{i})}.
\end{equation}

Substituting Eqs. (\ref{Residue02}) and (\ref{Residuei3}) into Eq. (\ref{Salphaofomegathermo2}) leads to
\begin{eqnarray}\label{Salphaofomegathermo3}
S_{\alpha}(\omega)&=&-\frac{\tilde{d}\tilde{a}}{a}\omega+\alpha\frac{\tilde{d}}{\sqrt{a}}\left(\frac{\tilde{a}b}{2a}-\tilde{b}\right)+\\
\nonumber &+&\alpha\delta_{\alpha,\text{sign}(\omega)}\begin{cases}
                  R(z_{1})+R(z_{2}), & \mbox{for } |\omega|<\omega_{1} \\
                  R(z_{2}), & \mbox{for } \omega_{1}<|\omega|<\omega_{2} \\
                  R(z_{2})+R(z_{3}), & \mbox{for } |\omega|>\omega_{3} \\
                  0, & \mbox{otherwise},
                \end{cases}
\end{eqnarray}

Inserting Eq. (\ref{Salphaofomegathermo3}) into Eqs. (\ref{Fofomega})-(\ref{Cofomega}) (with the corresponding substitutions of the parameters) yields for
\begin{equation}
\nonumber F(\omega),C(\omega)=4\omega-\sum_{\alpha}S_{\alpha}(\omega)
\end{equation}
the final results
\begin{eqnarray}\label{Fofomegafinal}
F(\omega)&=&\left(4+\frac{2a'}{a}\right)\omega\,-\\
\nonumber &-&\,\text{sign}(\omega)\,\begin{cases}
                  R(z_{1})+R(z_{2}), & \mbox{for } |\omega|<\omega_{1} \\
                  R(z_{2}), & \mbox{for } \omega_{1}<|\omega|<\omega_{2} \\
                  R(z_{2})+R(z_{3}), & \mbox{for } |\omega|>\omega_{3} \\
                  0, & \mbox{otherwise},
                \end{cases},
\end{eqnarray}
\begin{eqnarray}\label{Cofomegafinal}
C(\omega)&=&\left(4+\frac{\bar{a}}{a}\right)\omega\,-\\
\nonumber &-&\,\text{sign}(\omega)\,\begin{cases}
                  R(z_{1})+R(z_{2}), & \mbox{for } |\omega|<\omega_{1} \\
                  R(z_{2}), & \mbox{for } \omega_{1}<|\omega|<\omega_{2} \\
                  R(z_{2})+R(z_{3}), & \mbox{for } |\omega|>\omega_{3} \\
                  0, & \mbox{otherwise},
                \end{cases},
\end{eqnarray}
In the above we used that $\sum_{\alpha}1=2$, $\sum_{\alpha}\alpha=0$, and $\sum_{\alpha}\alpha\delta_{\alpha,\text{sign}(\omega)}=\text{sign}(\omega)$. Since $z_{i}(\omega)$ are even functions, $F(\omega)$ and $C(\omega)$ are odd functions.

Coming now to $g(k)=0.125|\bar{\bar{v}}(k)|^{2}=0.125|\bar{v}(k)|^{2}f^{\ast}(k)/|f(k)|$, using $2\sin k=-i(z-z^{-1})$ and $2\cos k=z+z^{-1}$, we write
\begin{equation}\label{fofz}
f(z)=\frac{M(z)}{z},
\end{equation}
where
\begin{equation}\label{Mofz}
M(z)=\frac{-i}{2}[(t+\Delta)z^{2}-\mu z+(t-\Delta)],
\end{equation}
and, hence,
\begin{equation}\label{fastovermodf}
\frac{f^{\ast}(z)}{|f(z)|}=\left[\frac{f^{\ast}(z)}{f(z)}\right]^{1/2}=z\left[\frac{M^{\ast}(z)}{M(z)}\right]^{1/2}.
\end{equation}

Recall from Eqs. (\ref{gofk}), (\ref{gofz}), (\ref{falphaofz}) and (\ref{Nofz}) that
\begin{equation}\label{vbarofz}
\frac{1}{4}|\bar{v}(k)|^{2}\rightarrow\frac{\bar{N}(z)}{z^{2}},
\end{equation}
where $\bar{N}(z)=\tilde{N}(z)$ with the ``tilded" parameters receiving the ``barred" ones.

Combining Eqs. (\ref{fastovermodf})-(\ref{vbarofz}) yields
\begin{equation}\label{gofz2}
g(z)=\frac{\bar{N}(z)}{2z}\left[\frac{M^{\ast}(z)}{M(z)}\right]^{1/2},
\end{equation}
and for $f_{\alpha}(z)=g(z)/D_{\alpha}(z)$
\begin{equation}\label{falphaofz2}
f_{\alpha}(z)=\frac{\bar{N}(z)M^{\ast^{1/2}}(z)}{2zM^{1/2}(z)D_{\alpha}(z)}.
\end{equation}

Since $\bar{N}(z)M^{\ast^{1/2}}(z)$ is holomorphic in $z$, the poles of $f_{\alpha}(z)$ are the zeros of $zD_{\alpha}(z)$. The zeros $z_{i}$ of $M^{1/2}(z)$ are also zeros of $M^{\ast^{1/2}}(z)$ and, hence, $M^{\ast^{1/2}}(z_{i})/M^{1/2}(z_{i})\rightarrow1$ so that these $z_{i}$ are not poles of $f_{\alpha}(z)$. Thus, $z_{0}=0$ and the roots $z_{i}$, $i=1,...,4$, of Eq. (\ref{poles}) are poles of multiplicity one.

Inputting into Eq. (\ref{Salphaofomegathermo}) the same conditions given in Eqs. (\ref{conds}) and the new multiplicity of $z_{0}$, we write
\begin{eqnarray}\label{Salphaofomegathermo2b}
S_{\alpha}(\omega)&=&\text{Res}[f_{\alpha},z_{0}]+\\
\nonumber &+&\begin{cases}
                  \text{Res}[f_{\alpha},z_{1}]+\text{Res}[f_{\alpha},z_{2}], & \mbox{for } |\omega|<\omega_{1} \\
                  \text{Res}[f_{\alpha},z_{2}], & \mbox{for } \omega_{1}<|\omega|<\omega_{2} \\
                  \text{Res}[f_{\alpha},z_{2}]+\text{Res}[f_{\alpha},z_{3}], & \mbox{for } |\omega|>\omega_{3} \\
                  0, & \mbox{otherwise},
                \end{cases}
\end{eqnarray}
with $z_{0},\,z_{1},\,z_{2},\,z_{3}$ of multiplicity 1.

Let us now evaluate the new residues. Employing Eq. (\ref{Residue}) with $n=1$ we get
\begin{equation}
\nonumber \text{Res}[f_{\alpha},z_{i}]=\lim_{z\rightarrow z_{i}}[(z-z_{i})f_{\alpha}(z)],
\end{equation}
which, inserting Eq. (\ref{falphaofz2}), yields
\begin{equation}\label{Residue0b}
\text{Res}[f_{\alpha},z_{0}]=\lim_{z\rightarrow 0}\left[\frac{\bar{N}(z)M^{\ast^{1/2}}(z)}{2M^{1/2}(z)D_{\alpha}(z)}\right]
\end{equation}
and
\begin{equation}\label{Residueib}
\text{Res}[f_{\alpha},z_{i}]=\lim_{z\rightarrow z_{i}}\left[(z-z_{i})\frac{\bar{N}(z)M^{\ast^{1/2}}(z)}{2zM^{1/2}(z)D_{\alpha}(z)}\right].
\end{equation}

Using the previously given limits and $\lim_{z\rightarrow0}M(z)=-i(t-\Delta)/2$, Eq. (\ref{Residue0b}) reduces to
\begin{equation}\label{Residue0b2}
\text{Res}[f_{\alpha},z_{0}]=\alpha\frac{i\bar{a}}{2\sqrt{a}}.
\end{equation}

Applying the L'Hopital rule, Eq. (\ref{Residueib}) becomes
\begin{equation}\label{Residueib2}
\text{Res}[f_{\alpha},z_{i}]=\frac{\bar{N}(z_{i})M^{\ast^{1/2}}(z_{i})}{d_{z}[2zM^{1/2}(z)D_{\alpha}(z)]_{z_{i}}},
\end{equation}
if $d_{z}[2zM^{1/2}(z)D_{\alpha}(z)]_{z_{i}}\neq0$.

Now
\begin{eqnarray}
\nonumber &d_{z}&[2zM^{1/2}(z)D_{\alpha}(z)]_{z_{i}}=\\
\nonumber &=&2M^{1/2}(z_{i})D_{\alpha}(z_{i})+2z_{i}\frac{M'(z_{i})}{2M^{1/2}(z_{i})}D_{\alpha}(z_{i})+\\
\nonumber &+&2z_{i}M^{1/2}(z_{i})D'_{\alpha}(z_{i})=
\end{eqnarray}
\begin{eqnarray}
\nonumber &=&2z_{i}M^{1/2}(z_{i})[\omega-\\
\nonumber &-&\frac{\alpha}{2N^{1/2}(z_{i})}(4az_{i}^{3}+3bz_{i}^{2}+2(2a+c)z_{i}+b)]=
\end{eqnarray}
\begin{eqnarray}
\nonumber &=&2M^{1/2}(z_{i})[\alpha N^{1/2}(z_{i})-\\
\nonumber &-&\frac{\alpha}{2N^{1/2}(z_{i})}(4az_{i}^{4}+3bz_{i}^{3}+2(2a+c)z_{i}^{2}+bz_{i})]=
\end{eqnarray}
\begin{equation}
\nonumber =\frac{\alpha2M^{1/2}(z_{i})}{2N^{1/2}(z_{i})}[2N(z_{i})-(4az_{i}^{4}+3bz_{i}^{3}+2(2a+c)z_{i}^{2}+bz_{i})],
\end{equation}
\begin{equation}
\nonumber d_{z}[2zM^{1/2}(z)D_{\alpha}(z)]_{z_{i}}=\frac{\alpha M^{1/2}(z_{i})}{N^{1/2}(z_{i})}(-2az_{i}^{4}-bz_{i}^{3}+bz_{i}+2a).
\end{equation}

Applying Eqs. (\ref{falphaofz}), (\ref{Dalphaofz}) and (\ref{fofz}) to obtain
\begin{equation}
\nonumber \frac{M^{1/2}(z)}{N^{1/2}(z)}=\frac{1}{zM^{\ast^{1/2}}(z)}
\end{equation}
and using Eq. (\ref{Lofz}), we arrive at
\begin{equation}\label{derivative2}
d_{z}[2zM^{1/2}(z)D_{\alpha}(z)]_{z_{i}}=\frac{\alpha L(z_{i})}{z_{i}M^{\ast^{1/2}}(z_{i})}.
\end{equation}

The zeros of Eq. (\ref{derivative2}), which are the same as those of Eq. (\ref{derivative}), do not contribute to Eq. (\ref{Salphaofomegathermo2b}).

Combining Eqs. (\ref{derivative2}) and (\ref{Residueib2}) and incorporating the constraint that $\alpha=+[-]$ implies $\omega>0[<0]$ wields
\begin{equation}\label{Residueib3}
\text{Res}[f_{\alpha},z_{i}]=\alpha\,\bar{R}(z_{i})\,\delta_{\alpha,\text{sign}(\omega)},
\end{equation}
where
\begin{equation}\label{Rbarzi}
\bar{R}(z_{i})=\frac{z_{i}\bar{N}(z_{i})M^{\ast}(z_{i})}{L(z_{i})},
\end{equation}
which, using Eqs. (\ref{gofz}) and (\ref{fofz}), reads
\begin{equation}\label{Rzi}
R(z_{i})=\frac{z_{i}^{2}g(z_{i})f^{\ast}(z_{i})}{L(z_{i})}.
\end{equation}

Substituting Eqs. (\ref{Residue0b2}) and (\ref{Residueib3}) into Eq. (\ref{Salphaofomegathermo2b}) leads to
\begin{eqnarray}\label{Salphaofomegathermo3b}
S_{\alpha}(\omega)&=&\alpha\frac{i\bar{a}}{2\sqrt{a}}+\\
\nonumber &+&\alpha\delta_{\alpha,\text{sign}(\omega)}\begin{cases}
                  \bar{R}(z_{1})+\bar{R}(z_{2}), & \mbox{for } |\omega|<\omega_{1} \\
                  \bar{R}(z_{2}), & \mbox{for } \omega_{1}<|\omega|<\omega_{2} \\
                  \bar{R}(z_{2})+\bar{R}(z_{3}), & \mbox{for } |\omega|>\omega_{3} \\
                  0, & \mbox{otherwise},
                \end{cases}
\end{eqnarray}

Inserting Eq. (\ref{Salphaofomegathermo3b}) into Eq. (\ref{Dofomega}) yields for
\begin{equation}
\nonumber D(\omega)=i\bar{\varepsilon}_{\text{e}}-\sum_{\alpha}\alpha S_{\alpha}(\omega)
\end{equation}
the final result
\begin{eqnarray}\label{Dofomegafinal}
D(\omega)&=&i\left(\bar{\varepsilon}_{\text{e}}-\frac{\bar{a}}{\sqrt{a}}\right)\,-\\
\nonumber &-&\,\begin{cases}
                  \bar{R}(z_{1})+\bar{R}(z_{2}), & \mbox{for } |\omega|<\omega_{1} \\
                  \bar{R}(z_{2}), & \mbox{for } \omega_{1}<|\omega|<\omega_{2} \\
                  \bar{R}(z_{2})+\bar{R}(z_{3}), & \mbox{for } |\omega|>\omega_{3} \\
                  0, & \mbox{otherwise},
                \end{cases}.
\end{eqnarray}
In the above, we used that $\alpha^{2}=1$, $\sum_{\alpha}1=2$, and $\sum_{\alpha}\delta_{\alpha,\text{sign}(\omega)}=1$. Since $z_{i}(\omega)$ are even functions so is $D(\omega)$.

\section{Analytical computation of functions entering the Green's functions for different regimes of the parameters}

Functions $G(\omega)$, $\bar{G}(\omega)$, and $\bar{\bar{G}}(\omega)$ are defined by:
\begin{equation}\label{Gofomega}
G(\omega)=-\sum_{k,\alpha}0.25|h(k)|^{2}\delta(\omega-\alpha|f(k)|),
\end{equation}
\begin{equation}\label{G1ofomega}
\bar{G}(\omega)=-\sum_{k,\alpha}0.125|\bar{v}(k)|^{2}\delta(\omega-\alpha|f(k)|),
\end{equation}
\begin{equation}\label{G2ofomega}
\bar{\bar{G}}(\omega)=-\sum_{k,\alpha}\alpha\,0.125|\bar{\bar{v}}(k)|^{2}\delta(\omega-\alpha|f(k)|).
\end{equation}

Let us define a generic function $g(\cos(k))=-0.25|h(k)|^{2},-0.125|\bar{v}(k)|^{2},-\alpha\,0.125|\bar{\bar{v}}(k)|^{2}$ which depends on $k$ through $\cos(k)$ only. Indeed, for $h(k)$ and $\bar{v}(k)$,
\begin{equation}
\nonumber g(k)=-\tilde{d}\frac{1}{4}[(-2\tilde{t}\cos k+\tilde{\mu})^{2}+(2\tilde{\Delta}\sin k)^{2}],
\end{equation}
\begin{equation}\label{gofkgeneric}
g(\cos(k))=-\tilde{d}[(\tilde{t}^{2}-\tilde{\Delta}^{2})\cos^{2}k-\tilde{t}\tilde{\mu}\cos k+\frac{\tilde{\mu}^{2}}{4}+\tilde{\Delta}^{2}],
\end{equation}
where $\tilde{d}=1\,[1/2]$ and $\tilde{x}=x'\,[\bar{x}]$, $x=t,\mu,\Delta$, when using $h(k)\,[\bar{v}(k)]$.

Meanwhile, since $f(k)=0.5[2\Delta\sin k+i(-2t\cos k+\mu)]$, $-\alpha0.125|\bar{\bar{v}}(k)|^{2}=-\alpha0.125|\bar{v}(k)|^{2}f^{\ast}(k)/|f(k)|=-\alpha0.125|\bar{v}(k)|^{2}[f^{\ast}(k)/f(k)]^{1/2}$ yields
\begin{eqnarray}
\nonumber g(k)=&-&\frac{\alpha}{8}[(-2\bar{t}\cos k+\bar{\mu})^{2}+(2\bar{\Delta}\sin k)^{2}]\times\\
\nonumber &\times&\left[\frac{2\Delta\sin k-i(-2t\cos k+\mu)}{2\Delta\sin k+i(-2t\cos k+\mu)}\right]^{1/2},
\end{eqnarray}
\begin{eqnarray}\label{gofkgenericbarbar}
\nonumber g(\cos(k))&=&-\frac{\alpha}{2}[(\bar{t}^{2}-\bar{\Delta}^{2})\cos^{2}k-\bar{t}\bar{\mu}\cos k+\frac{\bar{\mu}^{2}}{4}+\bar{\Delta}^{2}]\times\\
&\times&\left[\frac{2\Delta(1-\cos^{2}k)^{1/2}-i(-2t\cos k+\mu)}{2\Delta(1-\cos^{2}k)^{1/2}+i(-2t\cos k+\mu)}\right]^{1/2},
\end{eqnarray}
when using $\bar{\bar{v}}(k)$.

Eqs. (\ref{Gofomega})-(\ref{G2ofomega}) are all condensed in
\begin{equation}\label{Sofomega}
S(\omega)=\sum_{k,\alpha}g(\cos(k))\delta(\omega-\alpha|f(k)|),
\end{equation}
which, upon evaluating the sum over $k$, becomes
\begin{equation}\label{Sofomega2}
S(\omega)=\begin{cases}
            \sum_{k_{\alpha},\alpha}g(\cos(k_{\alpha}))\delta_{\alpha,\text{sign}(\omega)}, & \mbox{if } \exists\,k_{\alpha}\\
            0, & \mbox{otherwise}.
          \end{cases}
\end{equation}
where $k_{\alpha}\in[-\pi,\pi]$ are the solutions of $\omega=\alpha|f(k)|$. The factor $\delta_{\alpha,\text{sign}(\omega)}$ in Eq. (\ref{Sofomega2}) takes care of the fact that for solutions $k_{\alpha}$ to exist it is necessary (but not sufficient) that $\alpha=+$ for $\omega>0$ and $\alpha=-$ for $\omega<0$. If the constraint is satisfied, the sum over $k$ in Eq. (\ref{Sofomega}) yields the sum over the $k_{\alpha}$'s, if they exist, in Eq. (\ref{Sofomega2}), else it gives zero.

Evaluating the sum over $\alpha$ in Eq. (\ref{Sofomega2}) thus yields
\begin{equation}\label{Sofomega3}
S(\omega)=\begin{cases}
            \sum_{k_{\text{sign}(\omega)}}g(\cos(k_{\text{sign}(\omega)})), & \mbox{if } \exists\,k_{\text{sign}(\omega)}\\
            0, & \mbox{otherwise}.
          \end{cases}
\end{equation}

The solutions of $\omega=\alpha|f(k)|$, with $f(k)=0.5[2\Delta\sin k+i(-2t\cos k+\mu)]$, satisfy
\begin{equation}
\nonumber (t^{2}-\Delta^{2})\cos^{2}k-t\mu\cos k+\frac{\mu^{2}}{4}+\Delta^{2}-\omega^{2}=0,
\end{equation}
hence
\begin{equation}\label{kalphatau}
\cos(k)=c_{\pm}(\omega)=\frac{t\mu\pm\sqrt{D}}{2(t^{2}-\Delta^{2})},
\end{equation}
where $D=4(\Delta^{2}-t^{2})(\Delta^{2}-\omega^{2})+\Delta^{2}\mu^{2}$. Note that $c_{\pm}(\omega)$ are $\alpha$-independent and even functions of $\omega$.

Since $k$ is real, so is $\cos(k)$ and, thus, $D\geq0$ which, for $\Delta>t$, leads to $|\omega|\leq\omega_{3}=\Delta\{1+\mu^2/[4(\Delta^2-t^2)]\}^{1/2}$.

Also, the right-hand side of Eq. (\ref{kalphatau})
must be a real number between -1 and 1, hence $-2(\Delta^{2}-t^{2})+t\mu\leq\mp\sqrt{D}\leq2(\Delta^{2}-t^{2})+t\mu$, with $-\,[+]$ for $\tau=+\,[-]$. Taking all parameters to be positive, $2(\Delta^{2}-t^{2})+t\mu>0$, while $-2(\Delta^{2}-t^{2})+t\mu$ can be either positive or negative.

\vspace{.35cm}

$\bigstar$ Case $-2(\Delta^{2}-t^{2})+t\mu<0\,\Rightarrow\,\mu<2(\Delta^{2}-t^{2})/t$

\vspace{.35cm}

Combining $\mu<2(\Delta^{2}-t^{2})/t$ with the other constraints $\Delta>t$, $\mu<2t$ for $\mu>0$ (topological condition) yields that this regime of parameters is defined by $R_{1}(t,\mu,\Delta):\,\{0<\mu<\text{Min}\{2t,2(\Delta^{2}-t^{2})/t\},\Delta>t\}$.

Coming back to $-2(\Delta^{2}-t^{2})+t\mu\leq\mp\sqrt{D}\leq2(\Delta^{2}-t^{2})+t\mu$, with $-\,[+]$ for $\tau=+\,[-]$: If $\tau=-$, then $\sqrt{D}\leq2(\Delta^{2}-t^{2})+t\mu\,\Rightarrow\,|\omega|\geq t-\mu/2=\omega_{1}$. But if $\tau=+$, then $-2(\Delta^{2}-t^{2})+t\mu\leq-\sqrt{D}\,\Rightarrow\,|\omega|\geq t+\mu/2=\omega_{2}$.

Hence in the regime $R_{1}(t,\mu,\Delta)$, Eq. (\ref{kalphatau}) becomes
\begin{equation}\label{kalphatau2a}
c_{\tau}(\omega)=\begin{cases}
                          \frac{t\mu-\sqrt{D}}{2(t^{2}-\Delta^{2})}, & \mbox{if } \tau=- \mbox{ and } \omega_{1}\leq|\omega|\leq\omega_{3} \\
                          \frac{t\mu+\sqrt{D}}{2(t^{2}-\Delta^{2})}, & \mbox{if } \tau=+ \mbox{ and } \omega_{2}\leq|\omega|\leq\omega_{3} \\
                          \nexists, & \mbox{otherwise}.
                        \end{cases}
\end{equation}

Employing Eqs. (\ref{kalphatau}) and (\ref{kalphatau2a}) in Eq. (\ref{Sofomega3}) thus yields
\begin{equation}\label{Sofomega4a}
S(\omega)=\begin{cases}
            g(c_{-}(\omega)), & \mbox{if } \omega_{1}\leq|\omega|\leq\omega_{2}\\
            g(c_{-}(\omega))+g(c_{+}(\omega)), & \mbox{if } \omega_{2}\leq|\omega|\leq\omega_{3}\\
            0, & \mbox{otherwise}.
          \end{cases}
\end{equation}

\vspace{.35cm}

$\bigstar$ Case $-2(\Delta^{2}-t^{2})+t\mu>0\,\Rightarrow\,\mu>2(\Delta^{2}-t^{2})/t$

\vspace{.35cm}

Combining $\mu>2(\Delta^{2}-t^{2})/t$ with $\mu<2t$ for $\mu>0$ (topological condition) implies that $2(\Delta^{2}-t^{2})/t<2t\,\Rightarrow\,\Delta<\sqrt{2}t$. Adding the last constraint $\Delta>t$ thus wields that this regime of parameters is $R_{2}(t,\mu,\Delta):\,\{2(\Delta^{2}-t^{2})/t<\mu<2t,t<\Delta<\sqrt{2}t\}$.

Coming back to $-2(\Delta^{2}-t^{2})+t\mu\leq\mp\sqrt{D}\leq2(\Delta^{2}-t^{2})+t\mu$, with $-\,[+]$ for $\tau=+\,[-]$: If $\tau=-$, then $-2(\Delta^{2}-t^{2})+t\mu\leq\sqrt{D}\leq2(\Delta^{2}-t^{2})+t\mu\,\Rightarrow\,\omega_{1}\leq|\omega|\leq\omega_{2}$. Meanwhile, $\tau=+$ wields no solution.

Hence in the regime $R_{2}(t,\mu,\Delta)$, Eq. (\ref{kalphatau}) turns into
\begin{equation}\label{kalphatau2b}
c_{\tau}(\omega)=\begin{cases}
                          \frac{t\mu-\sqrt{D}}{2(t^{2}-\Delta^{2})}, & \mbox{if } \tau=- \mbox{ and } \omega_{1}\leq|\omega|\leq\text{Min}\{\omega_{2},\omega_{3}\}\\
                          \nexists, & \mbox{otherwise}.
                        \end{cases}
\end{equation}

Employing Eqs. (\ref{kalphatau}) and (\ref{kalphatau2b}) in Eq. (\ref{Sofomega3}) now gives
\begin{equation}\label{Sofomega4b}
S(\omega)=\begin{cases}
            g(c_{-}(\omega)), & \mbox{if } \omega_{1}\leq|\omega|\leq\text{Min}\{\omega_{2},\omega_{3}\}\\
            0, & \mbox{otherwise}.
          \end{cases}
\end{equation}

The functions in Eqs. (\ref{Gofomega})-(\ref{G2ofomega}) are obtained from Eq. (\ref{Sofomega4a}) [(\ref{Sofomega4b})] in the regime of parameters $R_{1}(t,\mu,\Delta)$ [$R_{2}(t,\mu,\Delta)$] by substituting the corresponding function $g$: Eq. (\ref{gofkgeneric}), with $\tilde{d}=1\,[1/2]$ and $\tilde{x}=x'\,[\bar{x}]$, $x=t,\mu,\Delta$, for Eq. (\ref{Gofomega}) [Eq. (\ref{G1ofomega})]; Eq. (\ref{gofkgenericbarbar}) for Eq. (\ref{G2ofomega}) where, in this case, the factor $\alpha$ in Eq. (\ref{gofkgenericbarbar}) yields that $S(\omega)$ comes with a multiplicative $\text{sign}(\omega)$. Since $c_{\tau}(\omega)$ are even functions, it follows that $G(\omega)$ and $\bar{G}(\omega)$ are even while $\bar{\bar{G}}(\omega)$ is odd.


\end{appendix}

\end{document}